%% file: Aalamifar_Lampe_arxiv.tex
\documentclass[journal,draftcls,onecolumn,12pt,twoside]{IEEEtranTCOM}
\usepackage{cite,url}
\usepackage{graphicx}
\usepackage{subfigure}
\usepackage{amsmath}
\usepackage{amsfonts}
\usepackage{amssymb}
\usepackage{color}
\usepackage{algorithm,algpseudocode}
\usepackage{epsfig}
\usepackage{epstopdf}
\usepackage{multirow}
\usepackage{graphicx}
\begin{document}

\begin{titlepage}
\title{{Cost-efficient QoS-Aware Data Acquisition Point Placement for Advanced Metering Infrastructure}}
\author { 
\IEEEauthorblockN{Fariba Aalamifar and Lutz Lampe} \\
\IEEEauthorblockA{Department of Electrical and Computer Engineering, The University of British Columbia, \ \{faribaa, lampe\}@ece.ubc.ca}
\fontsize{8}{10}
\thanks{This work has been submitted to the IEEE for possible publication. Copyright may be transferred without notice, after which this version may no longer be accessible}
}
\markboth{IEEE Transactions on Communications}
{Submitted paper}
\clearpage \maketitle
\thispagestyle{empty}
\vspace*{-2cm}

\begin{abstract}
\input{abstract_11Sept2016.tex}
\end{abstract}

\end{titlepage}

\graphicspath{{./figures/}}
\section{Introduction}\label{sec:intro}
As part of the broader concept of smart grids, advanced metering infrastructures (AMIs) are being massively deployed almost everywhere in the world. AMIs are responsible for reading the energy consumption from thousands of smart meters (SMs)~\cite{depuru2011smart,darby2006effectiveness,OpenSGForum,PAP2}, monitor the last-mile automated devices 
 for reporting emergency events~\cite{PAP2,OpenSGForum} such as electricity blackout and also unauthorized access to the power system~\cite{mclaughlin2009energy}. 
 In an AMI, due to the large number of devices and their distances, data collectors 
 are installed to collect traffic from several endpoints and transmit them to the utility control center on their behalf.

The placement of collector nodes, which are known as data acquisition points (DAPs)~\cite{PAP2,WiGrid} or aggregators~\cite{atkinson2014leveraging,bartoli2010secure} in smart grid communication networks (SGCNs) and as relay station~\cite{niyato2012cooperative}, gateway~\cite{aoun2006gateway} or sink~\cite{chatzigiannakis2006sink,sheng2006data} in broadband wireless access networks and sensor networks, respectively, has previously been  investigated~\cite{alsalih2010placement,lin2010optimal,lee2011eqar,lee2009qos}. 
However, there is a combination of features and requirements in AMI that render the problem sufficiently different from the data collector placement in other types of networks so that a new problem
formulation and solution for network planning are needed. %
For example, in sensor networks, the collector nodes can be placed on selected endpoint nodes~\cite{aoun2006gateway} or in arbitrary locations~\cite{alsalih2010placement}. Different from this, in a distribution grid with overhead powerlines, the utility poles are ideal locations for DAP placement~\cite{PAP2},  since this extends network coverage and also eliminates the cost of  new tower installations. 
Moreover, 
since the locations of utility poles are 
determined based on the power grid infrastructure, for example they are often located along roads and thus not uniformly distributed in a coverage area, it is not straightforward to apply the existing placement algorithms to place DAPs in AMIs. 
Another major difference  is that the on-time delivery of smart grid traffic to the utility control center and automated devices is critical for the correct operation of the electrical power grid~\cite{doe2010communications, OpenSGForum}. Also, due to the existence of two types of traffic classes namely, mission-critical and non-critical traffic, different scheduling schemes should be employed so that the quality of service (QoS) associated with both traffic can be maintained. In addition, due to the existence of rural areas, a multi-hop communication infrastructure is required in order to access further nodes.


Accordingly, the main design considerations for the placement of collector nodes in AMIs are the number and location of DAPs so that 1) the network coverage is ensured, 2) the required reliabilities associated with different types of smart grid traffic classes  are satisfied, and 3) existing infrastructures (utility poles) are used. 
Thereby, two types of access architectures from automated devices to DAPs are possible: a) direct and b) multi-hop communication. In this paper, we address the multi-hop connectivity case as it allows for accessing more remote devices and requires a smaller number of DAPs. 

The mathematical optimization formulation for DAP placement on top of existing utility poles is an integer programming (IP) problem and is NP-hard. 
For cases with small number of nodes, say no more than 200, the IBM CPLEX software~\cite{studio1software} and the GLPK solver~\cite{glpk} are typically used for finding optimized node locations. 
However, for cases with notably larger number of nodes, a heuristic algorithm needs to be developed~\cite{lin2010optimal,souza2013optimal,rolim2015modelling}. 


Heuristic algorithms proposed for relay placement are typically based on cover-set or facility-location algorithms. For example, references~\cite{aoun2006gateway} and~\cite{ali2011set}  propose weighted cover-set algorithms for respectively gateway and reader placement for wireless sensors and radio-frequency identification nodes. 
Reference~\cite{rolim2015modelling} applies the minimum-cover-set algorithm for finding the optimal location of DAPs for both single and multi-hop access in SGCNs. 
When the network becomes large, their heuristic algorithm breaks the area into smaller squares which can be handled by the optimizer. Their post-optimization step involves merging the solution of smaller squares by removing the redundant poles located in square edges. This step of their heuristic algorithm has a high complexity, because every pole that is not selected is checked to see if it can replace a subset of two or more selected poles. 
 In our preliminary work~\cite{aalamifar2014cost}, we have proposed a modified K-means algorithm for DAP placement in the single-hop communication scenario only considering network coverage, assuming SMs and poles are uniformly distributed through the area.
The K-means algorithm chooses random locations as primary potential locations for DAP placement and all the network construction is conducted based on these locations. These random locations are eventually mapped to the closest pole. However, there is a higher possibility that such a mapping would result in the violation of
QoS constraints when a realistic data set is considered, for example when poles are aligned with the road structure. 
Therefore, in this paper, we apply a different and more suitable heuristic algorithm by which the network is constructed from pole locations. 
In~\cite{souza2013optimal}, the authors develop a K-means based algorithm for placing a fixed number of aggregators on selected utility poles with the objective of minimizing the total number of hops SMs require to access the selected data aggregators. This work is among the first to consider multi-hop communication and minimize the experienced delays by minimizing the total number of hops.
However, limiting the number of hops only addresses the effect of transmission delay and ignores the effect of congestion delay which explicitly depends on the number of competitors and their arrival rates at each hop. 


%
References~\cite{NANkong2015} and~\cite{niyato2011machine} propose 
aggregator placement solutions for respectively 
maintaining and maximizing the obtained QoS in an AMI. They use M/D/1 and M/G/1 queuing models for computing the expected latency over the designed infrastructure. However, the mission-critical and non-critical smart grid traffic need the guarantees of certain latency requirements with certain probabilities (i.e., ensuring certain reliabilities), which is not provided through the solutions in~\cite{NANkong2015},~\cite{niyato2011machine}.

In this paper, we do not adopt the average latency model with fixed or minimum number of hops criteria considered in \cite{niyato2011machine,souza2013optimal,rolim2015modelling,NANkong2015}.  
Instead, to meet latency requirements of smart grid traffic, in Section~\ref{sec:sys}, we compute the probability of achieving a certain latency requirement for both mission-critical and non-critical traffic. To this end, we employ the IEEE 802.15.4g MAC protocol~\cite{802154gSUN2012} with the contention-free-period (CFP) and the contention-access-period (CAP) for scheduling critical and non-critical smart grid traffic.  
Then, we devise an optimization problem in Section~\ref{sec:sys} and propose a novel heuristic algorithm for solving the problem in Section~\ref{sec:alg}. 
The heuristic algorithm approximates the minimum required number of DAPs  through the use of a greedy algorithm for selecting potential pole locations for aggregator placement.  
In order to connect nodes through reliable routes, we use the Dijkstra algorithm for identifying transmission paths with the maximum packet success ratio. 
In Section~\ref{sec:res}, we provide performance results based on realistic locations for SMs and poles, which we have obtained from BC Hydro, a Canadian utility in the province of British Columbia. The results show that the paths found by our algorithm satisfy the latency requirements for both types of traffic to a specified level. We also compare the optimality and complexity of our solution for  small-scale scenarios with the branch and cut algorithm offered by the IBM CPLEX software~\cite{studio1software}. 
Finally, we conclude the paper in Section~\ref{sec:con}. 

\section{System Model and Problem Formulation}\label{sec:sys}
We consider a distribution grid with overhead power lines suspended from utility poles delivering electricity to homes
or businesses equipped with SMs. Some utility poles host DAPs, each of which is wirelessly connected to a
subset of the endpoints (SMs) either in a single-hop or multi-hop manner. The multi-hop communication 
utilizes 
IEEE 802.15.4g~\cite{802154gSUN2012} for connecting SMs  
to each other or to the DAPs. 
\renewcommand{\arraystretch}{1.2}
\begin{table}
\caption{Mission Critical and Non-Critical Traffic Properties~\cite{OSG}.}\label{tab:trafficProperties}
\vspace*{-13mm}
\begin{center}
\begin{tabular}{|c|c|c|c|c|c|c|}
\hline
\textbf{Traffic Class} & \textbf{Traffic Name}  & \textbf{Packet Size (Bytes)} & \textbf{Arrival Frequency}& \textbf{Traffic Type}& \textbf{Required Latency}  \\ 
\hline\hline
NC & Periodic Meter Reading (MR) & 250 & 15~min & Deterministic & 5~sec \\%
\hline
NC & On-demand  MR  Request & 50&  5~days &  Poisson &  30~sec \\ %
\hline
NC & On-demand  MR Response Data & 250  & 5~days & Poisson& 30~sec\\    %
\hline
MC & Power Quality  Notifications & 100  & 5~min & Poisson & 1~sec\\ 
\hline
MC & Remote Control Commands& 100& 1~day & Poisson & 1~sec\\%
\hline
MC & Alert Notifications & 50 & 1~week & Poisson & 3~sec\\%
\hline
\end{tabular}
\end{center}
\vspace*{-10mm}
\end{table}
We also assume that the following types of traffic, as listed in Table~\ref{tab:trafficProperties}, are passing through the grid.
\begin{enumerate}
\item Non-critical (NC) traffic such as reading the home energy consumption, periodically or on-demand.
\item Mission-critical (MC) traffic such as alert notifications, including meter tampering and power theft, remote control commands, and 
power quality 
(e.g., voltage, phase or current) notifications~\cite{mohassel2014survey}. The MC traffic is usually modelled according to a Poisson process~\cite{NANkong2015,GomezLeasedLine2013}.
\end{enumerate}

According to the OpenSG Forum~\cite{OpenSGForum}, reliability is defined as the probability that a packet can successfully be received at the destination within its required latency. Therefore, in order to meet the reliability requirements of the smart grid traffic, both the route quality in terms of the packet success rate    
 and the probability of exceeding the latency requirement over the route should be taken into account.
We formulate the link quality in Section~\ref{subsec:lq} and the probability of latency satisfaction for NC and MC traffic in Section~\ref{subsec:lat}. Using these expressions, we formulate the obtained reliability over a certain route in Section~\ref{subsec:rel}.


\subsection{Link Quality}\label{subsec:lq}
The link quality, defined as the probability of a successful packet transmission on the link between nodes $i$ and $j$, is obtained as
\begin{equation}\label{eq:PER}
1 - \epsilon_{ij} = 1 - \mathcal{Q} (\gamma_{ij}),
\end{equation}
where $\epsilon_{ij}$ is the link packet error rate (PER), $\gamma_{ij}$ is the signal to interference and noise ratio (SINR) 
and $\mathcal{Q}$ maps the SINR to the PER based on the modulation and coding scheme. The SINR is given by 
\begin{equation}\label{eq:SNR}
\gamma_{ij} = \frac {P_\mathrm{tx}}{(N_0' + I) \  PL(d_{ij}) \  \eta \   \ \delta}
\end{equation}
where $P_\mathrm{tx}$ is the transmit power, $PL$ is the distance-dependent path loss, the variable $d_{ij}$ denotes the distance between nodes $i$ and $j$, $\eta$ is the fading margin, and $N_0'= N_0 F$ where $N_0$ and $F$ are respectively the receiver noise power spectral density and noise factor. The variable $I$ denotes the interference%
, which accounts for the inter-operator interference when operating in the unlicensed band or when the same block of frequency is used by other operators or applications as well as cell-to-cell interference~\cite{PAP2014}. 
Furthermore, $\delta$ is the penetration loss which is present when SMs are located inside the building. 
The pathloss component $PL(d_{ij})$ depends on the area type. According to the NIST PAP2 guideline~\cite{PAP2014}, the Erceg SUI propagation model best emulates the channel propagation for rural and suburban scenarios. For urban areas, the ITU-R M.2135-1 (outdoor) and ITU-R M.1225 (indoor) propagation models are suggested. 

\subsection{Delay Model}\label{subsec:lat}
The IEEE 802.15.4g MAC protocol provides two types of medium access periods, namely CFP and CAP, within each frame. A node stores the MC and NC traffic in different queues, and schedules the mission-critical traffic through the CFPs using the time division multiple access (TDMA) scheme, and the non-critical traffic 
within the CAP time slots using the carrier sensing multiple access/collision avoidance (CSMA/CA) scheme.
We hereafter denote the number of available time slots per frame in the CFP and CAP by $N_{\mathrm{T}}$ and $N_{\mathrm{C}}$, respectively. 



Let us assume that the traffic from each node should be received at the destination within a time period of  $L$ seconds. 
In order to compute the probability that an NC or MC packet can be transmitted within this delay requirement, we need to translate $L$ to its equivalent number of available slots via 
\begin{eqnarray}\label{eq:availTimeslot}
N_{\mathrm{s}} = (\text{MC} \textrm{ or } \text{NC}) = \frac{L}{T_\mathrm{F}} \times N_{\mathrm{T}} \textrm{ or } N_{\mathrm{C}},  
\end{eqnarray}
where $T_\mathrm{F}$ is the frame duration in seconds. 
As we are dealing with a multihop communication system, the cumulative waiting time during all the hops should be less than the required latency.
Let us assume node $n$ is located at depth $H_n$ of the network and $r_{hn}$ is the relay node which forwards the message of node $n$ at hop $h$ where $ 1 \leq h \leq H_n$.
To meet the required delay for node~$n$ we allow
\begin{equation}\label{eq:AvailableSlots}
S = \left\lfloor{\frac{N_\mathrm{s}}{H_n}}\right\rfloor 
\end{equation}
time slots to be consumed at each of its forwarding nodes. 
This conservative assumption allows us to guarantee the required reliability. It should be noted that in practice, a larger delay may be consumed at some hop, while the total delay is still maintained. We hereafter assume each packet, even the largest-size packets of 250 bytes, can be transmitted within one time slot. 


There are several components included in the total packet delay, 
namely transmission, queuing, medium access, and propagation delay. 
Propagation delay is usually ignored for links with short distances~\cite{PAP2014}.
In the following, we first compute the average queuing delay. We then formulate the latency requirement that should be met for QoS satisfaction at each hop by deducting the queuing and transmission delay from the total allowed delay. Next, we mathematically derive the probability of meeting this required delay based on the MAC protocol specifications of the 802.15.4g standard.

\subsubsection{Queuing Delay}
For tractability of computing the queuing delay, we assume all sources generate Poisson traffic, which has been shown to be a sufficiently accurate approximation for mixed traffic as considered in our work~\cite{NANkong2015}.  We further assume that the Poisson traffic model also applies to nodes forwarding packets, which is justified if the traffic load at each node is low~\cite{shi2013analytical,Marco2012analytical,ray2005performance,shi2006starvation} and will also be verified numerically in Section~\ref{sim:validation} for typical traffic scenarios of our application.
We then accordingly apply the M/G/1 queueing model in order to compute  
the average waiting time at the queue of node $x$~\cite{shi2013analytical,Marco2012analytical}. According to the Pollaczek-Khinchin formula~\cite{bose2013introduction2q}, the waiting time~in time slots is given by
\begin{eqnarray}
T_{\mathrm{Q}_x} = \frac{\lambda_x E[Y_{x}^2]}{2 (1-\frac{\lambda_x}{\mu_x})} 
\end{eqnarray}
where
\begin{equation}\label{eq:lambda}
\lambda_x = \sigma_x \lambda_0 (N_{\mathrm{f}x} + 1)
\end{equation}
is the aggregated arrival rate at the node, $N_{\mathrm{f}x}$ denotes the total number of feeding nodes that are directly or indirectly connected to node $x$, $\lambda_0$ is the average traffic generation rate per node, and $\sigma_x$ gives the expected number of times that the packet should be re-transmitted, which will be calculated later in this section. 
$\mu_x$ is the packet service rate and 
$E[Y_{x}^2]$ denotes the second moment of the service time for both NC and MC traffic, which is given by
\begin{eqnarray}
E[Y_x^2] = \frac{N_\mathrm{C}+N_\mathrm{T}}{N_\mathrm{C}\  \textrm{or} \  N_\mathrm{T}}
\sum_{k=1}^{S} \Big(R_x(k)-R_x(k-1)\Big)  k^2,
\end{eqnarray}\label{eq:sp2}
 where $R_x(k)$ is the probability that the packet can successfully be transmitted within $k$ CAP or CFP slots. Variables $\mu_x$ and $R_x(k)$ are obtained later in this section. 



\subsubsection{Medium Access Delay}
Consider that $r_{hn}$ has $N_{r_{hn}}$ neighbours, which we collect in the set~$\Psi_{r_{hn}}$, and 
let $\mathcal{P}_{r_{hn}} = \{p_x: x \in \Psi_{r_{hn}} \}$ be the probabilities that these neighbours have a packet for transmission, given by 
$p_{x} = \frac{\lambda_{x}}{\mu_{x}}$~\cite{bose2013introduction2q}, 
where $\lambda_{x}$ has been defined in~\eqref{eq:lambda} above,
and $\mu_x$ is the service rate. $\mu_x$ is obtained later in the following section. 

Here, we describe how the probability of exceeding a certain delay is computed for the traffic generated by node $n$ for the above-mentioned scheduling schemes as a function of $S$ and $\mathcal{P}_{r_{hn}}$. In order to increase the obtained reliability, for each packet, we allow up to $N_{\mathrm{ARQ}}$ transmission attempts. 
\begin{itemize}

\item \textbf{Non-critical traffic:} 
Under the slotted CSMA/CA model, 
each node with the NC traffic, at each transmission attempt, would sense the channel at most $M+1$ times. At each sensing stage $m=0, 1, \cdots M$, it selects a random time slot within the backoff window, $W_m$, with equal probability. According to the IEEE~802.15.4g standard, in slotted CSMA/CA model, each node should identify the channel as idle for two consecutive slots before changing to transmission mode. 
If two nodes sense the channel as idle at the same time, there would be a collision. 
We note that 
since the length of CAP is comparable to the average size of a backoff stage, the accumulated traffic during CFP would be uniformly distributed over the CAP and therefore, similar to~\cite{Marco2012analytical,gao2009new,misic2006performance}, which consider inactive periods between CAPs, the probability that the channel is idle is assumed to be constant within the CAP.

\begin{figure}[!t]\centering{
\includegraphics[width=0.7\columnwidth]{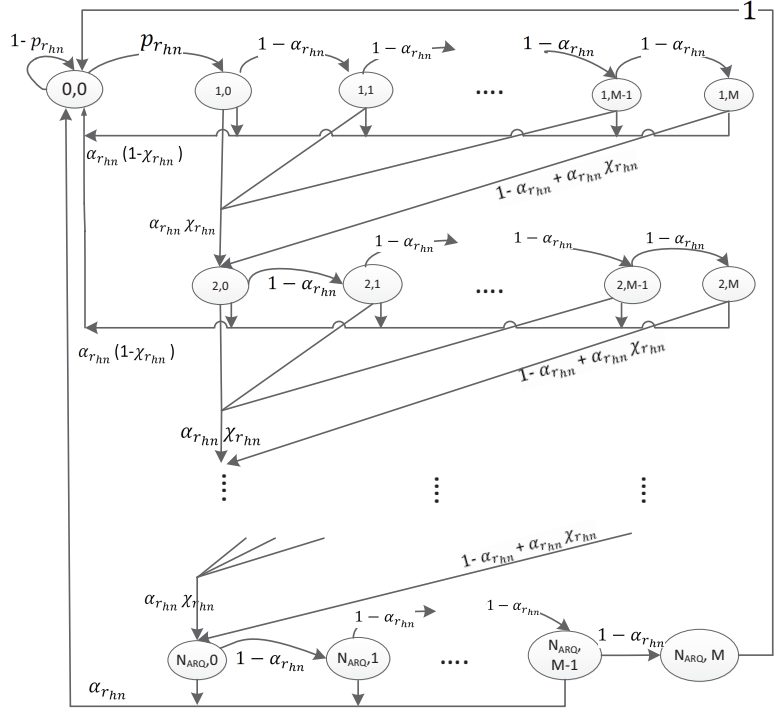}} 
\vspace*{-5mm}
\caption{Markov chain for the CSMA/CA process. State $(i,m),\  1 \leq i\leq N_{\mathrm{ARQ}}, \  0 \leq m \leq M$ represents the sensing stage $m$ in the $i$th transmission attempt, and $(0,0)$ is the state of having no packets for transmission. 
 $p_{r_{hn}}$ is the probability that the node has a packet for transmission, $\alpha_{r_{hn}}$ is the probability that the channel is idle and $1-\chi_{r_{hn}}$ is the probability that the packet has successfully been transmitted. 
} 
\vspace*{-9mm}
\label{fig:markov}
\end{figure}
Figure~\ref{fig:markov} shows the Markov chain model associated with the CSMA/CA procedure. 
We define $\beta_{1_{r_{hn}}}$ as the probability that the channel is busy when sensing for the first time, $\beta_{2_{r_{hn}}}$ as the probability that the channel is busy when sensing for the second time, provided that the channel was idle for the first time, and
\begin{equation}\label{eq:alpha}
\alpha_{r_{hn}} = (1-\beta_{1_{r_{hn}}}) (1-\beta_{2_{r_{hn}}})
\end{equation}
as the probability that the channel is determined as idle for two consecutive time slots. 
The channel is determined as busy if 
the channel was idle for two consecutive time slots and at least a node has sensed the channel in 
those slots. 
Hence, the probability of $\beta_{1_{r_{hn}}}$ is obtained from~\cite{gao2009new}
\begin{equation}
\beta_{1_{r_{hn}}} = (1-\beta_{1_{r_{hn}}}) (1-\beta_{2_{r_{hn}}}) \left(1-\left(\prod_{x \in \Psi_{r_{hn}}} (1 - \xi_x)\right)\right),
\end{equation}
where $\xi_x$ is the probability that a neighbour node conducts its first carrier sensing attempt in an arbitrary time slot. 
The probability that the channel is determined as busy when sensing for the second time, given that the channel was idle for the first time is obtained from~\cite{gao2009new}
\begin{eqnarray}
\beta_{2_{r_{hn}}} 
= (1-\beta_{2_{r_{hn}}}) \left(1-\left(\prod_{x \in \Psi_{r_{hn}}} (1 - \xi_x)\right)\right).
\end{eqnarray}

In order to compute $\xi_x$, we use the stationary probabilities associated with the Markov chain shown in Figure~\ref{fig:markov}.
Let $\boldsymbol{\pi}$ and $\mathbf{T}$ respectively denote the stationary distribution vector and transition matrix of this Markov chain. Solving the stationary state equation $\boldsymbol{\pi} \mathbf{T} = \boldsymbol{\pi}$ subject to $\sum_{j} \boldsymbol{\pi}_{j} = 1$, 
we can compute the probability of conducting the first carrier sensing attempt in an arbitrary time slot by a neighbour node as
\begin{equation}~\label{eq:xi}
\xi_{x} = \displaystyle \sum_{i=1}^{N_{\mathrm{ARQ}}} \displaystyle \sum_{m=0}^{M} \frac{\boldsymbol{\pi}_{g(i,m)}}{W_m},
\end{equation}
where $g(i,m) = (i-1)(M+1) + m + 1$, $\boldsymbol{\pi}_{g(i,m)}$ is the probability of being in sensing stage~$m$ in transmission attempt~$i$, and $\frac{1}{W_m}$ gives the probability of conducting the first carrier sensing attempt in an arbitrary time slot in stage $m$.

In order to compute the probability that a node can transmit its packet within the required latency, we need to compute the probability that the node senses the channel within the latency and also the channel is idle. Let us define $\theta_{r_{hn}}(k)$ as the probability that node~${r_{hn}}$ senses the channel in time slot~$k$ and also the channel is idle. 
Since slot~$k$ can be sensed at any of the $N_{\mathrm{ARQ}}$ transmission attempts and $M+1$ backoff stages, $\theta_{r_{hn}}(k)$ is obtained as 
\begin{eqnarray}\label{eq:WSP}
\theta_{r_{hn}}(k)  = \sum_{i = 1}^{N_{\mathrm{ARQ}}}  \sum_{m=0}^{M} \zeta_{r_{hn}}(k,i,m) \  \alpha_{r_{hn}},
\end{eqnarray}
where $\zeta_{r_{hn}}(k,i,m)$ is the probability of sensing the channel at slot $k$, in sensing stage~$m$, in transmission attempt~$i$. 
The variable $\zeta_{r_{hn}}(k,i,m)$  is computed based on the probability of having an unsuccessful transmission attempt (due to either finding the channel as busy during all $M+1$ backoff stages or due to packet transmission failure) in one of the previous $d$ slots in the previous try 
and then sensing the channel  at slot $k-d-2$ in the current try,
\begin{eqnarray}\label{eq:Pr_sense}
\zeta_{r_{hn}}(k,i,m) = \left\lbrace
\renewcommand{\arraystretch}{1.2}
\begin{array}{l}
\sum_{d=3(i-2)+1}^{k-2} \sum_{m' = 0}^{M}  \zeta_{r_{hn}}(d,i-1,m') \ 
 \Delta_{m'}\ \phi(k-d-2,m), \  i>1,  \\ 
\phi(k,m),  \quad i =1,  \\
\end{array} \right.
\end{eqnarray}
where at least 3~slots are consumed at each attempt (2~slots for sensing and 1 for transmission), 
\begin{eqnarray}
\Delta_{m'} = \left\lbrace
\renewcommand{\arraystretch}{1.2}
\begin{array}{l}
\alpha_{r_{hn}}  \chi_{r_{hn}},   \quad \quad \quad \quad \quad \quad \quad \ \ m' < M, \\
\alpha_{r_{hn}} \chi_{r_{hn}} + (1 - \alpha_{r_{hn}}),   \quad \quad m' = M,  
\end{array} \right.  \nonumber
\end{eqnarray}
and  
 $\phi(k,m)$ is the probability of assessing the channel at slot~$k$ in sensing stage~$m$. 
The value of $\phi(k,m)$ is also recursively computed as a cumulative probability of sensing the channel at slot $j$ in the previous sensing stage, finding the channel as busy in either the first or second slot and accordingly, backing off for $k-j$ slots with probability~$\frac{1}{W_m}$ in the current sensing stage~$m$~\cite{baz2015analysis}. In other words, $\phi(k,m)$ can be calculated as 
\begin{eqnarray}\label{eq:p_sensing_in_stage}
\phi(k,m) = 
\left\lbrace
\renewcommand{\arraystretch}{1.2}
\begin{array}{l}
\displaystyle \sum_{j=1}^{k-1} \phi(j,m-1) \  \beta_1 \ \frac{1}{W_m} 
+ \displaystyle \sum_{j=1}^{k-2} \phi(j,m-1) \ (1-\beta_1) \ \beta_2 \ \frac{1}{W_m}, \ \ m \geq 1, \  k \geq 1, \\
\frac{1}{W_0}, \quad m = 0,\quad k \geq 1,\\
0, \quad \quad \quad \quad \quad \quad k < 1.
\end{array} \right.  
\end{eqnarray}

Finally using~\eqref{eq:WSP} with~\eqref{eq:Pr_sense}-\eqref{eq:p_sensing_in_stage} and~\eqref{eq:alpha}-\eqref{eq:xi}, the probability that node~${r_{hn}}$ can successfully transmit the packet within the required latency is obtained as
\begin{eqnarray}\label{eq:NCLatency}
R_{r_{hn}}(S) = \sum_{k=1}^{S-T_{\mathrm{Q}}-1} \theta_{r_{hn}}(k) (1-\chi_{r_{hn}}),
\end{eqnarray}
where 
$1-\chi_{r_{hn}}$ is the probability that the packet can successfully be transmitted, i.e., the packet transmission does not fail due to a  collision (given that the channel is determined as idle, at least one other node senses the channel at the same time as $r_{hn}$) or due to a link error. It is obtained as
\begin{equation}\label{eq:pr_succ}
1-\chi_{r_{hn}} = (1-\epsilon_h) \left(\prod_{x \in \Psi_{r_{hn}}} (1 - \xi_x)\right),
\end{equation}
where $\epsilon_h$ is the link PER between $r_{hn}$ and the relay node at the next hop as defined in~\eqref{eq:PER}.
\item \textbf{Mission-critical traffic:}
In this section, we compute the probability that all the bandwidth requests from the neighbour nodes, can be scheduled within the latency requirement. 
According to~\cite{PAP2014}, this probability is computed as
\begin{eqnarray}\label{eq:poisscdf}
\Pr (\ell_{r_{hn}} \le  S) 
 = \sum_{i=0}^{S-1} \left( \sum_{\mathcal{\psi} \in {\Psi}_{r_{hn},i}} \ \ \  \prod_{j \in \cal{\psi}} p_j \prod_{k \in {\Psi}_{r_{hn}} \setminus \mathcal{\psi}} (1-p_k) \right),
\end{eqnarray}
where $\ell_{r_{hn}}$ is the experienced delay at relay node $r_{hn}$ over one transmission attempt, ${\Psi}_{r_{hn},i}$ is the set of all subsets of $\Psi_{r_{hn}}$ with size $i$. For Poisson traffic assumed here, the expression in~\eqref{eq:poisscdf} has the  closed-form solution~\cite{Fernandez_closedform}
\begin{equation}\label{eq:DFT_closedform}
\Pr (\ell_{r_{hn}} \le S) = 
\sum_{i=0}^{S-1} \frac{1}{N_{r_{hn}}+1} 
\displaystyle \sum_{\kappa=0}^{N_{r_{hn}}}  \mathrm{e}^{\mathrm{j}\frac{-2 \pi \kappa i}{N_{r_{hn}}+1}} \prod_{k=1}^{N_{r_{hn}}} \left(p_k \mathrm{e}^{\mathrm{j} \frac{2 \pi \kappa}{N_{r_{hn}}+1}} + (1-p_k)\right),
\end{equation}
where $\mathrm{j}$ is the imaginary unit. Let us define $\mathcal{L}_{{r_{hn}},i}$ as the cumulative sum of delays over $i$ transmission attempts. We can compute the obtained reliability at hop~$h$ after $N_{\mathrm{ARQ}}$ transmission attempts as
\begin{equation}\label{eq:relARQ}
R_{r_{hn}}(S) =  \displaystyle \sum_{i = 1}^{N_{\mathrm{ARQ}}} \Pr (\mathcal{L}_{{r_{hn}},i} \le S-T_{\mathrm{Q}} ) \ (\epsilon_h)^{i - 1} \ (1-\epsilon_h),
\end{equation}
where similar to the NC traffic, the probability of latency satisfaction at each attempt can be recursively computed based on the time that has elapsed in the previous attempts, i.e.,
\begin{equation}
\Pr (\mathcal{L}_{{r_{hn}},i} \le S) =  \displaystyle \sum_{k = i-1}^{S-1} \Pr (\mathcal{L}_{{r_{hn}},i-1} = k) \Pr (\ell_{r_{hn}} \le S - k), i > 1,
\end{equation}
where
\begin{equation}
\Pr (\mathcal{L}_{{r_{hn}},i} = k) = \displaystyle \sum_{d = i-1}^{k-1} \Pr (\mathcal{L}_{{r_{hn}},i-1} = d) \Pr (\ell_{r_{hn}} = k-d), i > 1,
\end{equation}
and
\begin{align}
\Pr (\ell_{r_{hn}} = u) = 
\displaystyle \frac{1}{N_{r_{hn}}+1}\sum_{\kappa=0}^{N_{r_{hn}}} \mathrm{e}^{\mathrm{j}\frac{-2  \pi \kappa (u-1)}{N_{r_{hn}}+1}} \prod_{k=1}^{N_{r_{hn}}} \left(p_k \mathrm{e}^{\mathrm{j}\frac{2 \pi \kappa}{N_{r_{hn}}+1}} + (1-p_k)\right).
\end{align}

\item{\textbf{Computing service rates:}}
As mentioned earlier, in order to compute $p_x$, we need to compute the average service rate for the NC and MC traffic for node $x$. The average service rate for node $x$ can be obtained as~$\mu_x = \frac{1}{E[Y_x]}$, where $E[Y_x]$ is the mean packet service time, 
which is calculated as 
\begin{IEEEeqnarray}{rCr}\label{eq:NCmu}
E[Y_x] =  
\left\lbrace
\renewcommand{\arraystretch}{1.3}
\begin{array}{l}
\displaystyle\frac{1}{N_\mathrm{C}+N_\mathrm{T}} \sum_{i=1}^{N_\mathrm{T}} i
+ \displaystyle \sum_{m=0}^{M} (1 - \alpha_x)^{m} 
\frac{W_m+2}{2} + \displaystyle \sum_{m=1}^{M} (1 - \alpha_x)^{m} \left(\frac{W_m+2}{2 N_C} N_\mathrm{T} \right) + 1, \quad \textrm{CSMA/CA}, \\\IEEEnonumber
\displaystyle\frac{1}{N_\mathrm{C}+N_\mathrm{T}} \sum_{i=1}^{N_\mathrm{C}} i + 
\displaystyle{\left\lfloor\frac{\displaystyle\frac{1}{2}\sum_{x' \in \Psi_x \cup x} \lambda_{x'} \frac{L}{H_x}}{N_\mathrm{T}}\right\rfloor \left(N_\mathrm{T}+N_\mathrm{C}\right)} +\!\!\!\!\!\! \mod(\frac{1}{2}\displaystyle\sum_{x' \in \Psi_x \cup x} \lambda_{x'} \frac{L}{H_x}, N_\mathrm{T}), \textrm{TDMA},
\end{array}\right. \IEEEyesnumber
\end{IEEEeqnarray}%
that is, for the NC traffic 
$E[Y_x]$ is computed based on whether the packet has arrived during the CFP and accordingly, the corresponding CFP duration should be added to the service time. Also, we need to consider the expected time that is needed for backoff, plus adding 
another CFP 
if the channel is busy and the remaining CAP slots are not sufficient for a new backoff. Finally, one time slot is added 
for packet transmission. 
For the MC traffic, $E[Y_x]$ is computed based on whether the packet has arrived during the CAP and accordingly, the corresponding CAP duration should be added to the service time. 
Also, we need to consider the expected CFP time that is required for serving packets that have been generated by the node and neighbours during the time period~$\frac{L}{H_x}$.


\item{\textbf{Computing expected number of retransmissions:}}
The value of $\sigma_{x}$ gives the expected number of retransmissions that is required for a successful transmission of a packet generated by node~$x$, which is located at hop~$h$. This value is obtained as~\cite{Bertsekas1992DataNet}
\begin{eqnarray}
\sigma_{x} = \left\lbrace
\renewcommand{\arraystretch}{1.2}
\begin{array}{l}
\frac{1}{1 - \epsilon_h},   \quad \quad \quad \quad \quad \quad \quad \quad \textrm{for MC traffic}, \\
\frac{1}{(1 - \chi_{x})(1 - (1-\alpha_{x})^{M+1})},   \quad \quad \textrm{for NC traffic}.  
\end{array} \right.  \nonumber
\end{eqnarray}
\end{itemize}

\subsection{Obtained Reliability over the Path}\label{subsec:rel}
Based on the derivation of reliability for each hop in~\eqref{eq:NCLatency} and~\eqref{eq:poisscdf}, 
the obtained reliability over each path can be calculated as
\begin{equation}\label{eq:delRetx}
R_n = \prod_{h=1}^{H_n}R_{r_{hn}}(S).
\end{equation}

\subsection{Problem Formulation}\label{sec:formulation}
In order to collect the traffic from SMs either in a single-hop or multi-hop structure, aggregators are placed on top of the existing utility poles. 
The placement should be conducted such that coverage for all automated devices is ensured, the required latency for critical and non-critical traffic is satisfied, and at the same time, a cost-efficient infrastructure in terms of installation and maintenance is obtained. 

To formulate the associated optimization problem 
let us 
assume $N_{\textrm{SM}}$ is the number of SMs in the area which need to be covered and $N_{\textrm{poles}}$ is the number of poles from which a subset should be selected for DAP placement. The binary variable $x_j$ indicates whether a DAP is installed on pole $j$. Also let the binary variables $y_{ij}$, $q_{ii'}$ and $z_{ii'}$  indicate whether an SM $i$ is directly connected to the DAP located on pole $j$, 
whether a node $i'$ is the immediate parent\footnote{Any node which is on the route from the source to the destination is defined as the ancestor of the source. The ancestor node directly connected to the source is called the source's parent.} of another node $i$, and whether node $i'$ is an ancestor of another node $i$, respectively.  
Using these variables and the expressions from Sections~\ref{subsec:lq} to \ref{subsec:rel}, we can write the optimization problem for the DAP placement in~\eqref{eq:optAll} (on the next page). 
According to~\cite{rolim2015modelling,aalamifar2014cost}, DAPs are very costly to be installed. 
Therefore, in order to have a cost-efficient infrastructure, we define the objective~\eqref{eq:opt} as the minimization of the
installation cost, $c_{\textrm{inst}}$, which we consider linearly proportional to the total number of DAPs that should be mounted on top of the poles.
Assuming that discovering one route is enough for each SM, constraint~\eqref{eq:singleormultihop} ensures that it is either directly connected to a DAP or it has an immediate connection to another smart meter, which becomes its parent node. Constraint~\eqref{eq:ParRel} provides the relation between the parent of a node, $q_{ii'}$, and its ancestors, $z_{ii'}$. Constraints~\eqref{eq:loop} and~\eqref{eq:loop1} ensure that only one of the nodes $i$ or $i'$ can be the parent or an ancestor of the other one. 
Constraint~\eqref{eq:coverage} enforces the connectivity of all nodes to a DAP, via single or multi-hop communication. 
Accordingly, constraint~\eqref{eq:probExcDelay} as previously obtained in~\eqref{eq:delRetx}, ensures the satisfaction of the reliability constraint as a cumulative effect of packet success ratio and the latency requirement for both MC and NC traffic, where $\rho$ is the specified required reliability in percentage. Constraint~\eqref{eq:cap} ensures that the aggregated traffic from the connected nodes to each DAP is less than the offered service rate by the DAP, $\mu$. 
Constraint~\eqref{eq:CR} ensures that the relation between DAP selection and placement is maintained, i.e., an SM can only be connected to a pole which is selected for DAP installation.

\begin{figure*}[!t]
\vspace{-7mm}
\begin{subequations}\label{eq:optAll}
\begin{eqnarray}
\hspace*{-1cm}\min_{\begin{array}{ccc}
\{x_j\},\{y_{ij}\},\{q_{ii'}\},
\{z_{ii'}\}
\end{array}
} \ \ c_{\textrm{inst}} =\displaystyle\sum_{j=1}^{N_{\textrm{poles}}} x_j & &\label{eq:opt}\\ 
 \textrm{Subject to }  \quad \quad \quad \quad \quad \quad \quad \quad \quad \quad    & &\quad \nonumber  \\
\sum_{j=1}^{N_{\textrm{poles}}}y_{ij}+\sum_{i'=1}^{N_{\textrm{SM}}} q_{ii'} = 1 ,&\quad 1 \leq i \leq N_{\textrm{SM}},& \label{eq:singleormultihop}\\ 
q_{ii'} \leq z_{ii'}, &\quad  1 \leq i, i' \leq N_{\textrm{SM}},& \label{eq:ParRel} \\
q_{ii'} + q_{i'i} \leq 1,&\quad  1 \leq i, i' \leq N_{\textrm{SM}},& \label{eq:loop}\\
z_{ii'} + z_{i'i}	\leq 1,&\quad  1 \leq i, i' \leq N_{\textrm{SM}},& \label{eq:loop1}\\
\sum_{j=1}^{N_{\textrm{poles}}} y_{ij}+\sum_{j=1}^{N_{\textrm{poles}}}\sum_{i'=1}^{N_{\textrm{SM}}}z_{ii'} y_{i'j} = 1,&\quad  1 \leq i \leq N_{\textrm{SM}},& \label{eq:coverage}\\
R_{i} \geq \rho , &\quad 1 \leq i \leq N_{\textrm{SM}}, \textrm{ for MC and NC,}&\label{eq:probExcDelay} \\
\sum_{i=1}^{N_{\textrm{SM}}} y_{ij}\lambda_{i} \leq \mu, & \quad 1 \leq j \leq N_{\textrm{poles}},&\label{eq:cap} \\
y_{ij} \leq x_j,&\quad 1 \leq i \leq N_{\textrm{SM}},\quad 1\leq j \leq N_{\textrm{poles}},& \label{eq:CR} \\
x_j, y_{ij}, z_{ii'}, q_{ii'}  \in \{0,1\}, &\quad  1 \leq i, i' \leq N_{\textrm{SM}}, \quad 1\leq j \leq N_{\textrm{poles}},& \label{eq:binary}
\end{eqnarray}
\end{subequations}
\hrulefill
\vspace{-7mm}
\end{figure*}
\vspace{-7mm}
\begin{figure*}[!t]\center
\subfigure[]{\includegraphics[width=0.25\paperwidth]{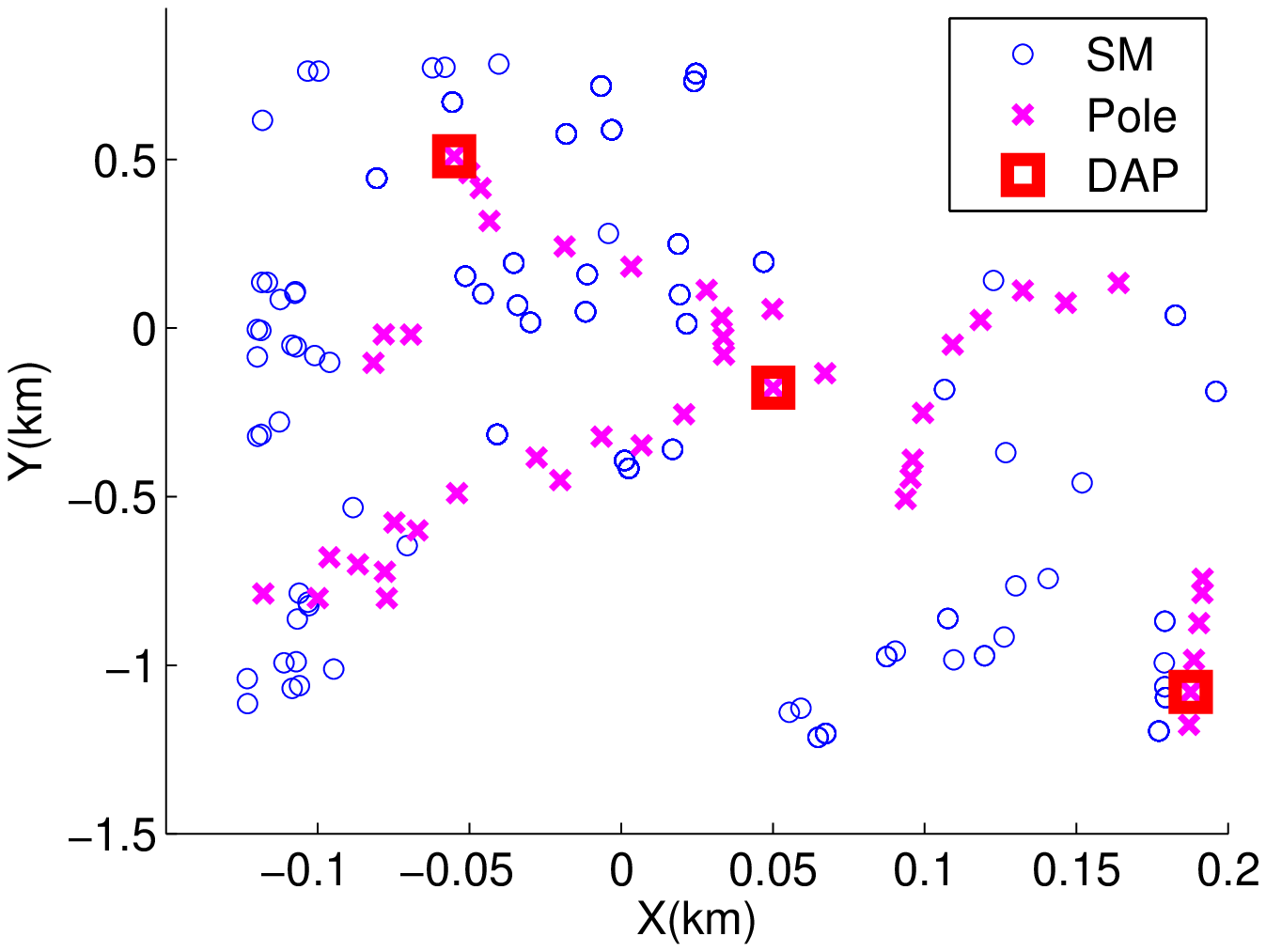}\label{fig:firstphase}}
\subfigure[]{\includegraphics[width=0.25\paperwidth]{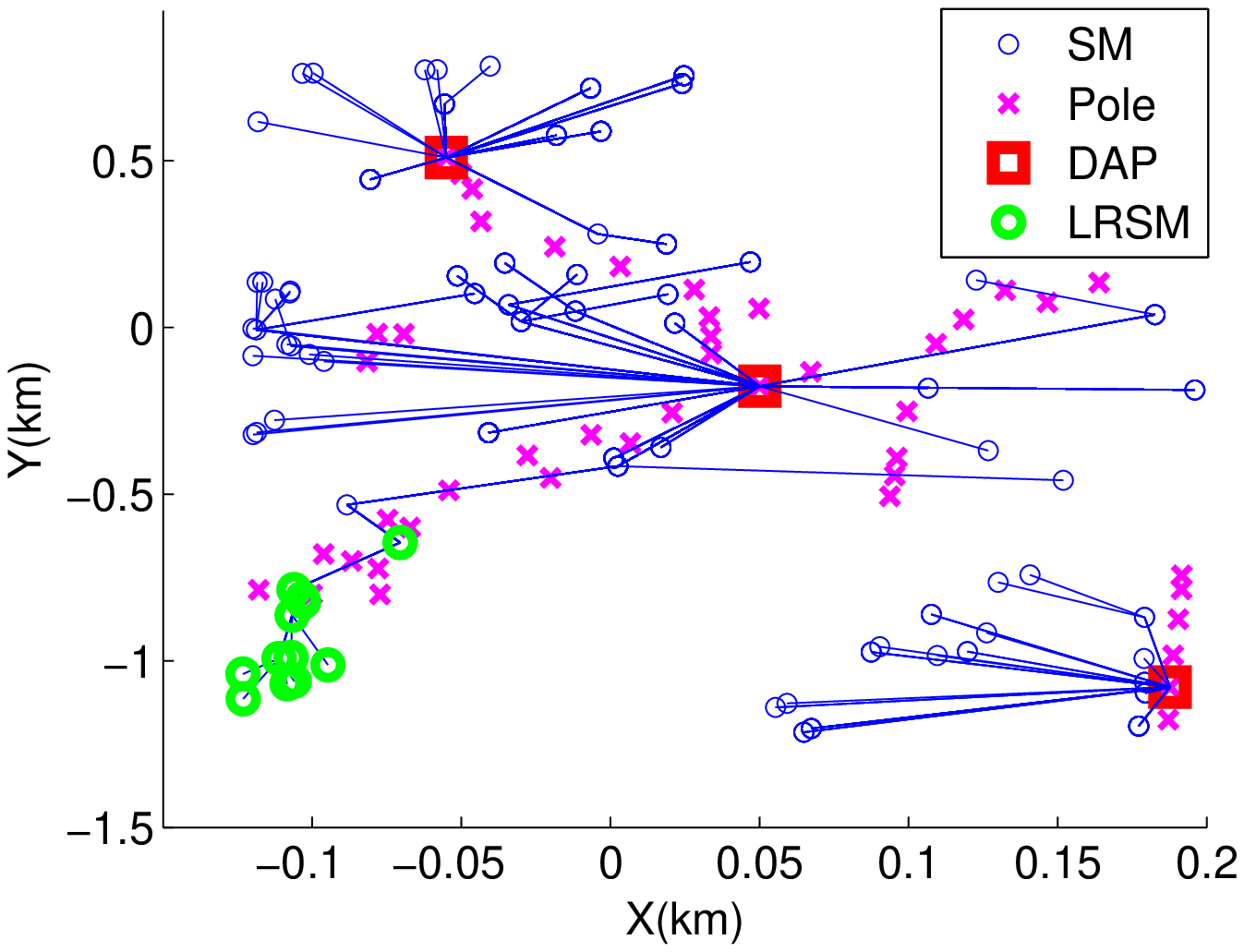}\label{fig:secondphase}}
\subfigure[]{\includegraphics[width=0.25\paperwidth]{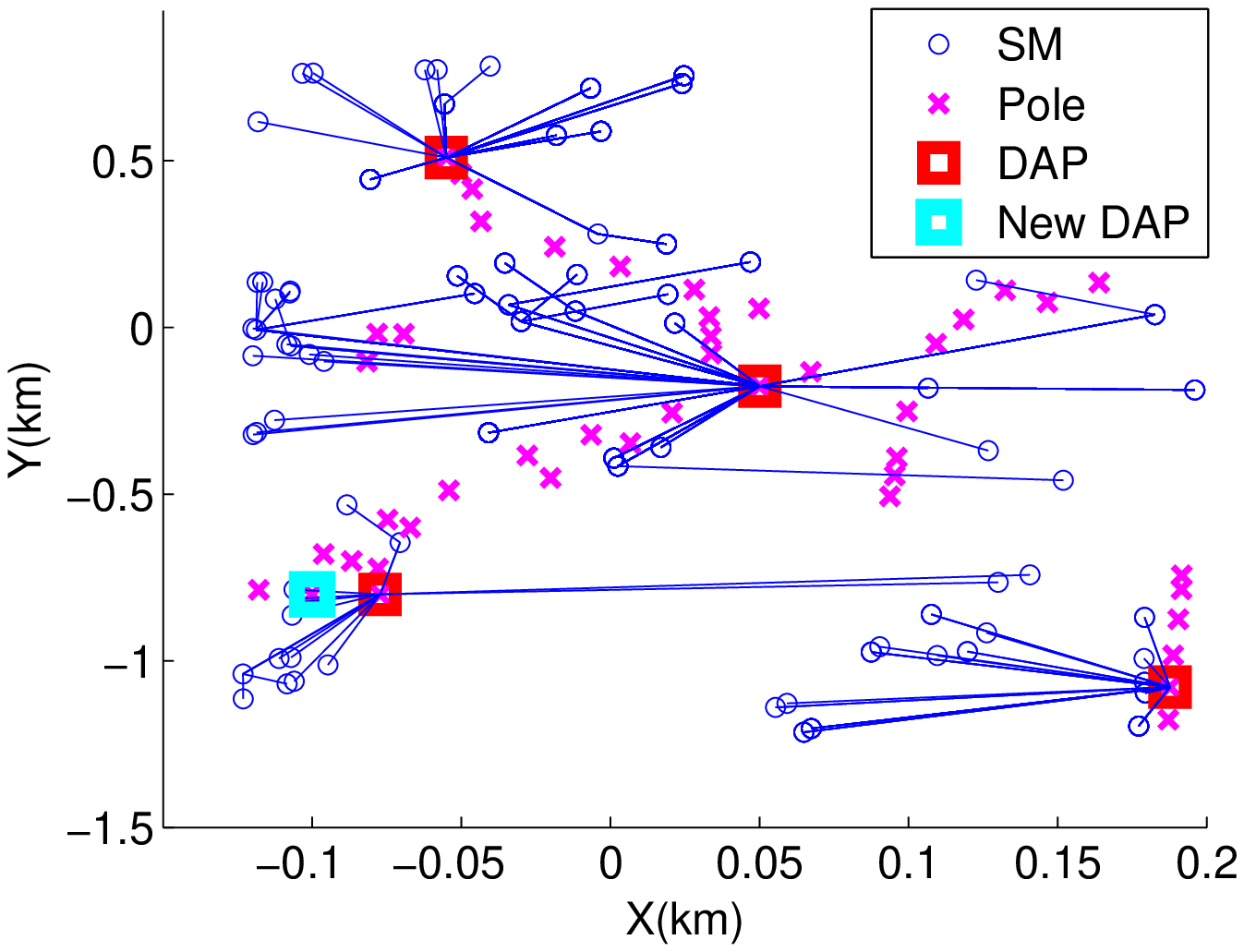}\label{fig:secondphase_switch_poles}}
\caption{Sample scenario for illustration of the steps of the heuristic algorithm. (a) First phase pole selection, (b) Second phase - step I, with initial shortest paths (LRSM denotes an SM which experiences low reliability) and second phase - step II (DAP locations have not been changed in this case), (c) Second phase - step III, placing a new DAP at $(-0.1, -0.8)$ and re-running second phase - step I for re-constructing the tree, and second phase - step II, relocating each aggregator closer to the center-point of its current cluster members (the new aggregator is moved to $(-0.07,-0.8)$).}\label{fig:example_phases}
\vspace{-12mm}
\end{figure*}

\vspace{7mm}
\section{DAP Placement Algorithm}\label{sec:alg}
The optimization in \eqref{eq:optAll} is an IP problem and directly solving it has 
an exponential time complexity with regards to the problem size, i.e., number of variables and constraints~\cite{meng2014smart}. Optimization solvers such as CPLEX~\cite{studio1software} and GLPK~\cite{glpk} employ the branch and cut method for solving IP problems. However, the complexity of such algorithms is still high and  exponential in the worst case scenario. Therefore for large networks, a lower complexity algorithm is desired~\cite{lin2010optimal,alsalih2008placement,lee2011eqar}. 
In this section, we propose a new heuristic algorithm, which is partly inspired from~\cite{aoun2006gateway} and~\cite{lin2010optimal}, and uses a greedy approach for identifying potential locations for relay placement.
We later on, through the results presented in Section~\ref{sec:res}, show that our proposed algorithm can provide a good solution to the DAP placement problem with a relatively low computational complexity. 

The proposed DAP placement algorithm consists of two phases. In the first phase, we address the objective~\eqref{eq:opt} through approximating the minimum required number of aggregators and their initial locations. This is done through selecting poles that cover the largest number of uncovered SMs through multi-hop communication as per~\eqref{eq:singleormultihop}-\eqref{eq:coverage}. In the second phase, 
based on the initial location of DAPs, we explore shortest path routes for the SMs to connect them to the DAPs and ensure that their network coverage, and QoS and capacity requirements as per~\eqref{eq:probExcDelay} and~\eqref{eq:cap} are maintained.

\subsection{Phase~1: Pole Selection}\label{Sec:AlgPoleSelection}



In this phase, through a greedy approach, we select the poles that have the largest number of connectivities to the uncovered SMs as candidates for DAP installation. In order to identify the set of SMs that can be covered by a certain pole through multi-hop communication  as per~\eqref{eq:singleormultihop}-\eqref{eq:coverage},
we construct a $k$-dimensional (KD) tree\footnote{A KD tree is a data structure for organizing $k$-dimentional data points in a binary search tree~\cite{samet1990design}. Performing range search operation over this tree (data structure) helps to identify the set of nodes that are in the communication range of certain locations.} over the set of SMs and perform range search operations, considering the effective coverage range of poles and SMs, $d_{\textrm{smax}}$ and $d_{\textrm{pmax}}$.

We repeat the above step for the remaining SMs that are not yet connected to a selected pole until all SMs are connected to a DAP or there is no solution for the remaining nodes, i.e. there is no pole or SM in their communication range. 

\vspace*{-4mm}
\subsection{Phase~2: Tree Construction} 
In this phase, we connect endpoints to the aggregators that have been selected in phase 1 and ensure that the capacity and QoS requirements~\eqref{eq:probExcDelay} and \eqref{eq:cap} are satisfied. 
We perform the following steps.

\textbf{Step~I (route discovery):} We use the Dijkstra algorithm to connect each SM through single or multi-hop communication, to the DAP that its capacity has not yet exceeded as per~\eqref{eq:cap} and also results in obtaining the maximum packet success rate. 
To this end, we use the link PERs obtained from~\eqref{eq:PER} and~\eqref{eq:SNR} via
\[
c_{ij} = \log\left(\frac{1}{1-\epsilon_{ij}}\right)
\]
as the link costs. This step determines the clusters, i.e., the set of SMs that are connected to each DAP. 

\textbf{Step~II (relocating each aggregator to the center-point of its cluster members):} 
As the first phase of the algorithm only addresses the coverage constraint, in this phase 
 we move each DAP to the pole nearest to the center-point of its cluster members, so that on average fewer hops would be required for SMs within the cluster to access the DAP and accordingly, a better reliability can be provided for them. 
Note that all the SMs should be able to connect to the newly selected location for the DAP, otherwise, this re-location would not be conducted.


\textbf{Step~III (adding new aggregators):} In this step, we compute the obtained reliability as per~\eqref{eq:probExcDelay} for all the nodes and disconnect those that experience low reliability for either of their MC or NC traffic. 
Then, we re-run the first phase of the algorithm for finding new aggregators for covering the disconnected nodes. As there might be some already connected nodes whose reliability would improve if they connected to the newly added aggregators, we repeat the second phase of the algorithm over the whole set of SMs in order to re-connect them to the new set of DAPs. 
Adding new aggregators can only increase satisfaction of the reliability constraint, and thus this step is re-iterated until the required reliability is met for all nodes 
or  no solution can be found (i.e., no solution exists for meeting the required reliability).

Figure~\ref{fig:example_phases} shows an example of the phases of our algorithm in an SGCN with 425 SMs and 45 poles. The smart meters are 
shown as circles, poles are marked with crosses and the selected DAPs are represented as squares. 
As it can be seen, the first phase of the algorithm selects three poles for DAP installation (Figure~\ref{fig:firstphase}). The second phase of the algorithm constructs initial shortest paths for all the nodes and computes their obtained packet success ratio and reliability. We can observe that 13 nodes become disconnected during step~III of phase~2 (marked as larger (green) circles in Figure~\ref{fig:secondphase}) as their obtained reliability with the current set of DAPs is less than the specified reliability of $\rho=~98\%$.
 Then, through repeating the first phase of the algorithm, a new pole 
 is selected for the DAP placement (new DAP in Figure~\ref{fig:secondphase_switch_poles}) and steps~I and~II of the second phase are repeated for reconstructing the shortest paths and moving poles to the center-point of their currently allocated cluster members.

In this section, we provide details on the performance of the proposed algorithm in terms of optimality 
and convergence speed.

\subsubsection{Optimality Analysis}
The DAP placement is an instance of the set cover problem~\cite[Theorem1]{chandra2004optimizing} and we have applied a greedy approach for solving it. It is well-known that the approximation factor of greedy algorithms for solving a set cover problem in the worst-case scenario is $\ln(N)$, where $N$ is the number of nodes to be covered~\cite{chandra2004optimizing}.
Moreover, there is no approximation algorithm that can provide a significantly better approximation factor than what is provided by a greedy algorithm for solving a set cover problem~\cite{Feige1998SetCover}. Therefore, the solution provided by the proposed heuristic algorithm in the worst case differs from the optimal solution by a factor of $\ln(N_{\mathrm{SM}})$, and this is the best approximation factor that a  polynomial solution can achieve.

\subsubsection{Convergence Analysis}
According to the global convergence theorem, an algorithm converges to a desired solution if we can define a descent function on the solution set~\cite{zangwill1969nonlinear}.
Since in each iteration of our algorithm, the number of nodes that are not covered by a DAP are decreasing (adding new DAPs improves the experienced reliability)
, we can conclude that our algorithm converges.

In terms of the convergence ratio, assume $r_k$ is the number of DAPs in the $k$th iteration of the algorithm, and $r^*$ is the number of DAPs when the algorithm converges. 
Since in our algorithm,
$\nu = \lim_{k \rightarrow \infty} \frac {r_{k+1} - r*}{r_k - r^*} $ is a value between $0$ and $1$ (as the distance to the required number of DAPs is decreasing), according to~\cite{luenberger1973introduction} we can conclude that the algorithm linearly converges to the desired solution with ratio $\nu$. The value of $\nu$ is different for different scenarios. 
For a smaller value of $\nu$, the algorithm converges faster.

\section{Numerical Results and Discussion}\label{sec:res}
\input{simulation_R3_9Dec2017_arxiv.tex} 

\vspace*{-4mm}
\subsection{Complexity Analysis}\label{subsec:complexity}
\subsubsection{Proposed algorithm}
Here we estimate the complexity of each step in our algorithm to derive its overall complexity.

\textbf{KD tree construction and range search:}
In the first phase of our placement algorithm, we use the KD tree data structure for storing SM locations. Then, we perform a range search operation over this tree in order to identify the set of SMs which are in the communication range of a certain pole. The runtime and memory complexity of KD tree construction are respectively $O(N_{\mathrm{SM}} \log({N_{\mathrm{SM}}}))$  and $O(N_{\mathrm{SM}})$. The range search operation complexity is $O(N_{\mathrm{poles}} \log(N_{\mathrm{SM}}))$.




\textbf{Shortest path:} In order to identify optimal routes for each SM, shortest paths are constructed from each DAP using the Dijkstra algorithm. The associated time and memory complexity are 
 $O(N_{\mathrm{SM}}^2)$ and $O(N_{\mathrm{SM}}^2)$, respectively.

Since the shortest-path search has the higher complexity of the above two steps, the total algorithm run-time and memory complexities are of the orders of $N_{\mathrm{DAP}} O(N_{\mathrm{SM}}^2)$ and $N_{\mathrm{DAP}} O(N_{\mathrm{SM}}^2)$, respectively. For the specific Kamloops scenario with 8053 SMs considered above, we measured a memory usage of 83~MB.


\subsubsection{CPLEX}\label{sec:cplex}
CPLEX uses a branch and cut algorithm for finding the optimal solution to the IP problem. 
In the worst case, the complexity of such an algorithm is exponential, and the actual mean time-complexity depends on many factors and is evaluated empirically~\cite{CPLEXHelp,ResearchGate}.
Another limiting factor when optimization solvers are used for solving IP problems is the required RAM. According to~\cite{CPLEXForum}, for every 1000 constraints, at least 1 MB RAM is required by CPLEX in order to solve an IP problem. Since the presented DAP problem in~\eqref{eq:optAll} considering 8053 SMs and 776 poles has around 140,000,000 constraints, an estimated 140~GB RAM would be needed to solve it by CPLEX.


\subsection{Comparison with Other Works}
In this section, we compare the optimality and time-complexity of our algorithm with the work presented in~\cite{rolim2015modelling} and~\cite{niyato2011machine}. 
For a fair comparison with~\cite{rolim2015modelling}, we limit the number of hops to $H=4$ and compare the solution of our algorithm with the second scenario in~\cite[Table II]{rolim2015modelling} that has a similar number of SMs and poles as the Kamloops scenario.
We observe from our simulations results, which are omitted here due to space constraints, that our algorithm finds a more cost-efficient solution as it only selects 37 out of 776 poles and ensures coverage and latency constraints, while the algorithm from~\cite{rolim2015modelling} selects 426 poles for DAP placement and only ensures SM coverage. Furthermore, the complexity of their 
algorithm is higher. In particular, the method from~\cite{rolim2015modelling} requires to calculate the multi-hop connectivity matrix as part of the pre-processing method, 
which has a computational complexity of $H \cdot O((N_{\textrm{poles}} + N_{\textrm{SM}})^3)$. 
Then, the coverage matrix is passed to the GLPK software for obtaining the minimum number of cover sets, which in the worst-case scenario has a complexity of~$O(2^{(N_{\textrm{poles}} + N_{\textrm{SM}})})$. When the network becomes large, their heuristic algorithm breaks the area into smaller squares which can be handled by the optimizer. Their post-optimization step involves merging the solution of smaller squares, solved by GLPK software, by removing the redundant poles located in square edges. This step has the complexity of $O(N_{\textrm{SM}} N_{\textrm{poles}}^2)$. In terms of the memory complexity, the method from~\cite{rolim2015modelling} would require 2-306~MB depending on the selected square size.


Reference~\cite{niyato2011machine} utilizes the divide and conquer algorithm for identifying the set of SMs that can relay traffic in an AMI. In the procedure of relay selection, the maximization of QoS is considered in the objective by minimizing packet loss and average latency, which are calculated based on the link distance and M/D/1/k queueing theory. The algorithm focuses on single-hop connectivity of endpoints to the aggregator and finally, connects every 10-15 endpoints to one aggregator. This is not a feasible solution in practice, since at least around 533 aggregators would then need to be installed and maintained.



%
%
%

\section{Conclusion}\label{sec:con}
In this paper, the problem of DAP placement for an AMI with overhead power lines has been investigated. We proposed a mutli-phase heuristic algorithm for selecting the optimized pole locations for DAP placement such that smart grid QoS requirements can be met.
We maximize the obtained reliability for the smart grid traffic through discovering routes with minimum packet error rates and scheduling the mission-critical and the non-critical traffic using  TDMA and CSMA/CA protocols, respectively. The probability of exceeding a certain latency is computed based on the specific characteristics of these two protocols. 
Comparing the results of our algorithm with the literature and solutions obtained by the IBM CPLEX software for small-scale examples, we believe that our algorithm is competitive in terms of performance for the problem at hand, albeit at much lower complexity. The complexity advantage allows us to successfully tackle
larger-scale problems as shown in this paper. 

\section{Acknowledgment}
This research has been supported by the Natural Sciences and Engineering
Research Council (NSERC) of Canada. The authors would
like to thank BC Hydro for their valuable help in providing us with the required data set.

\footnotesize
\bibliographystyle{IEEEtran}
\bibliography{IEEEabrv,references_abrv}

\end{document}

%% file: abstract_11Sept2016.tex
In an advanced metering infrastructure (AMI), data acquisition points (DAPs) are responsible for collecting traffic from several smart meters and automated devices  and transmitting them to the utility control center. 
 Although the problem of optimized data collector placement has already been addressed for wireless broadband and sensor networks, DAP placement is quite a new research area for  AMIs. 
In this paper, we investigate the minimum required number of DAPs and their optimized locations on top of the existing utility poles in a distribution grid 
such that smart grid quality of service requirements can best be provided. 
In order to solve the problem for large-scale AMIs, 
we devise a novel heuristic algorithm using a 
 greedy approach for identifying potential pole locations for DAP placement and the Dijkstra's shortest path algorithm for constructing reliable routes. 
We employ the characteristics of medium access schemes from the IEEE 802.15.4g smart utility network (SUN) standard, and consider mission-critical and non-critical smart grid traffic. The performance and time-complexity of our algorithm are compared with those obtained by the IBM CPLEX software for small scenarios. 
Finally, we apply our devised DAP placement algorithm to examples of realistic smart grid AMI  topologies. 

%% file: simulation_R3_9Dec2017_arxiv.tex
In this section, we test our proposed DAP placement algorithm using realistic smart meter and pole locations information from the area of Kamloops, BC, Canada.





\subsection{Simulation Settings}
\begin{figure*}[!t] \centering{
\subfigure[]{\includegraphics[width=0.75\paperwidth]{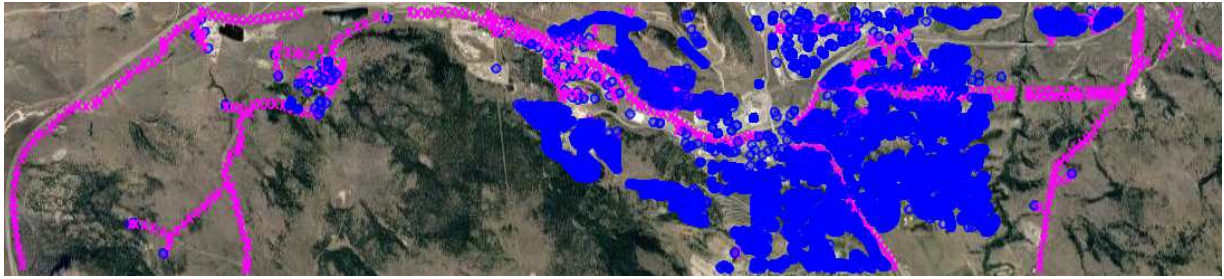}\label{fig:gmap}}
\hspace*{-8mm}
\subfigure[]{\includegraphics[width=0.75\paperwidth]{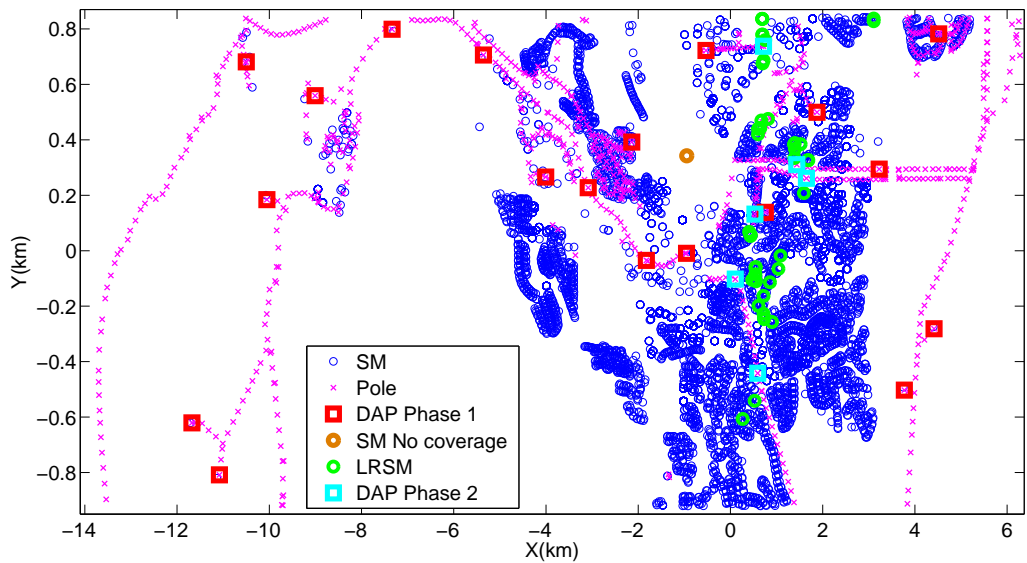}
\label{fig:daptopology}}} 
\caption{(a) The geographic location of smart meters and poles in the Kamloops suburban area. 
(b) Results of the proposed DAP placement algorithm for the Kamloops scenario.
The red and cyan squares show the poles that are selected for DAP placement respectively in the first and second phase of the algorithm. The green circles show the low-reliability SMs (LRSMs) for which the poles in the second phase were added. 
The larger (orange) circle identifies the 19~SMs that are not connected to any DAP.
}
\label{fig:DAP_placement}
\vspace*{-10mm}
\end{figure*}

Table~\ref{tab:sim} summarizes the parameters we have used for running our simulations. 
Figure~\ref{fig:gmap}
presents the geographical locations of SMs and poles over the map of Kamloops, BC, Canada. The SMs and poles are marked with blue circles and magenta crosses, respectively. It is important to note that the poles are mostly aligned with the roads on the map and their location do not follow a uniform-random distribution model that is sometimes assumed in the literature.
As suggested in~\cite{PAP2}, the Erceg Type~B best models the signal propagation for the smart grid infrastructure in rural and suburban areas. Therefore, we have used this model for emulating the pathloss in the considered Kamloops suburban area, which is a hilly environment with light to moderate number of trees. The area size is $20 \times 2~\textrm{km}^2$ which includes 8053 SMs and 776 poles. 
The traffic specifications are derived from~\cite{OpenSGForum} as presented in Table~\ref{tab:trafficProperties} in Section~\ref{sec:sys}.

\renewcommand{\arraystretch}{0.9} 
\begin{table}
\caption{Simulation parameters~\cite{PAP2,OSG}.}\label{tab:sim}
\vspace*{-14mm}
\center
\begin{tabular}{|c|c|||c|c|||c|c|}
\hline
\textbf{Parameter} & \textbf{Value} & \textbf{Parameter} & \textbf{Value} & Parameter & Value  \\
\hline\hline   
Req. reliability, $\rho$ & 90\% & PL model  &   Erceg Type B & Bandwidth & $281$~kHz (802.15.4g)  \\%
\hline
$N_{\mathrm{ARQ}}$ & 4 & Interference Margin ($I_\textrm{m}$) & 6 dB & Fading Margin ($\eta$) & $12.3$~dB \\
\hline
SM / DAP height & $2$ / $10$~m & Transmission power ($P_{\textrm{tx}}$) & $30$~mW & Modulation and  &  QPSK  \\\cline{1-4} 
Noise Factor ($F$) & 7~dB & Receiver Noise PSD ($N_0$) & $-174$~dBm/Hz & coding scheme (MCS) & code rate of $\frac{3}{4}$  \\  
\hline
\end{tabular}
\vspace*{-7mm}
\end{table}

\vspace{-7mm}
\subsection{Performance Comparison with CPLEX}
We first compare the optimality and complexity of our devised algorithm with the results obtained based on the CPLEX software for solving~\eqref{eq:optAll}. To this end, since CPLEX is not able to solve the large-scale scenarios, we select smaller scale scenarios 
considering different area densities from the Kamloops scenario. The performances of our algorithm and the CPLEX software are compared in Table~\ref{tab:cplexComp}. As the number of aggregators indicates the optimization objective, we can observe that our algorithm returns near-optimal results and at the same time, our algorithm offers much lower run-time complexity and memory requirement. 
We further observe from Table~\ref{tab:cplexComp} that more aggregators are required for the scenarios with lower SM density.
\begin{table*}
\renewcommand{\arraystretch}{0.67} 
\caption{Comparing the optimality and complexity of proposed DAP placement algorithm and CPLEX for solving problem~\eqref{eq:optAll}}
\vspace*{-8mm}
\label{tab:cplexComp}
\hspace*{-1cm}
\begin{tabular}{|c|c|c|c|c|c|c|}
\hline
Scenario & Method  & Memory (MB) & Time (sec.) & Number of & Number of  & Max.\\
         &         &                   &              & Iterations & Aggregators &    hops      \\
\hline
\hline
47 SMs   &          &  358.2              &             &         &           &      \\ 
43 Poles &  CPLEX   &  4487 Variables   &     25.0      &    NA     &  4        &   2   \\
Rural (23.5 SMs per km$^2$) & (13009 Non-zero coeffs.)  &  6746 Constraints    &   &       &          &        \\
\hline
47 SMs   &             &         &        5.0           &          &                       &   \\ 
43 Poles & DAP placement algorithm        &  0.7     &    4.3 First phase    & 1   &  4                    &  2      \\
Rural (23.5 SMs per km$^2$)   &  &    &    0.7 Second phase         &           &            &      \\ 
\hline
\hline
60 SMs   &          &  481.1              &                  &    &           &      \\ 
12 Poles &  CPLEX   &  4124 Variables   &     77.0           &  NA  &  1        &   10			   \\
Suburban (155.2 SMs per km$^2$) &  (12942 Non-zero coeffs.)  & 7841 Constraints    &   &       &          &        \\
\hline
60 SMs   &              &         &        6.1                &     &                       &   \\ 
12 Poles & DAP placement algorithm        &  0.9     &    5.1 First phase  & 2      &  2                    &  6      \\
Suburban (155.2 SMs per km$^2$)   &  &    &    1.0 Second phase         &     &                  &      \\ 
\hline
\hline
74 SMs   &   & 1094.6    & 1860.0       &      &   &       \\ 
37 Poles &   CPLEX     & 9554 Variables &    (Stopped at     &   NA & 1   &  5     \\
Suburban (513.9 SMs per km$^2$) & (34290 Non-zero coeffs.) & 15140 Constraints  &  $6\%$ optimality gap) &       &    &        \\ 
\hline
74 SMs   &   &            &  7.3                    &              &         &       \\ 
37 Poles &   DAP placement algorithm      & 1.2  & 6.5 First phase    &  1  &  1 &  5    \\
Suburban (513.9 SMs per km$^2$) & &   &    0.8 Second phase &         &              &        \\ 
\hline
\hline
161 SMs   &   & 854.3     &                       &       &                &       \\ 
24 Poles &    CPLEX     & 38117 Variables      &     840.0  &  NA  &   1               &  6     \\
Urban (958.3 SMs per km$^2$) & (135888 Non-zero coeffs.) &  64335 Constraints  &  &        &    &        \\ 
\hline
161 SMs   &   &            &  14.3                    &             &          &       \\ 
24 Poles &  DAP placement algorithm      &      2.3           &    10.3 First phase &  1      &  1                    &  6     \\
Urban (958.3 SMs per km$^2$) & &   &    4.0 Second phase         &                   &    &        \\ 
\hline
\end{tabular}
\hspace*{-1cm}
\end{table*}



\vspace*{-8mm}
\subsection{Validation of the Delay Model}\label{sim:validation}
\begin{figure}[!t] 
\subfigure[]{\includegraphics[width=0.37\paperwidth]{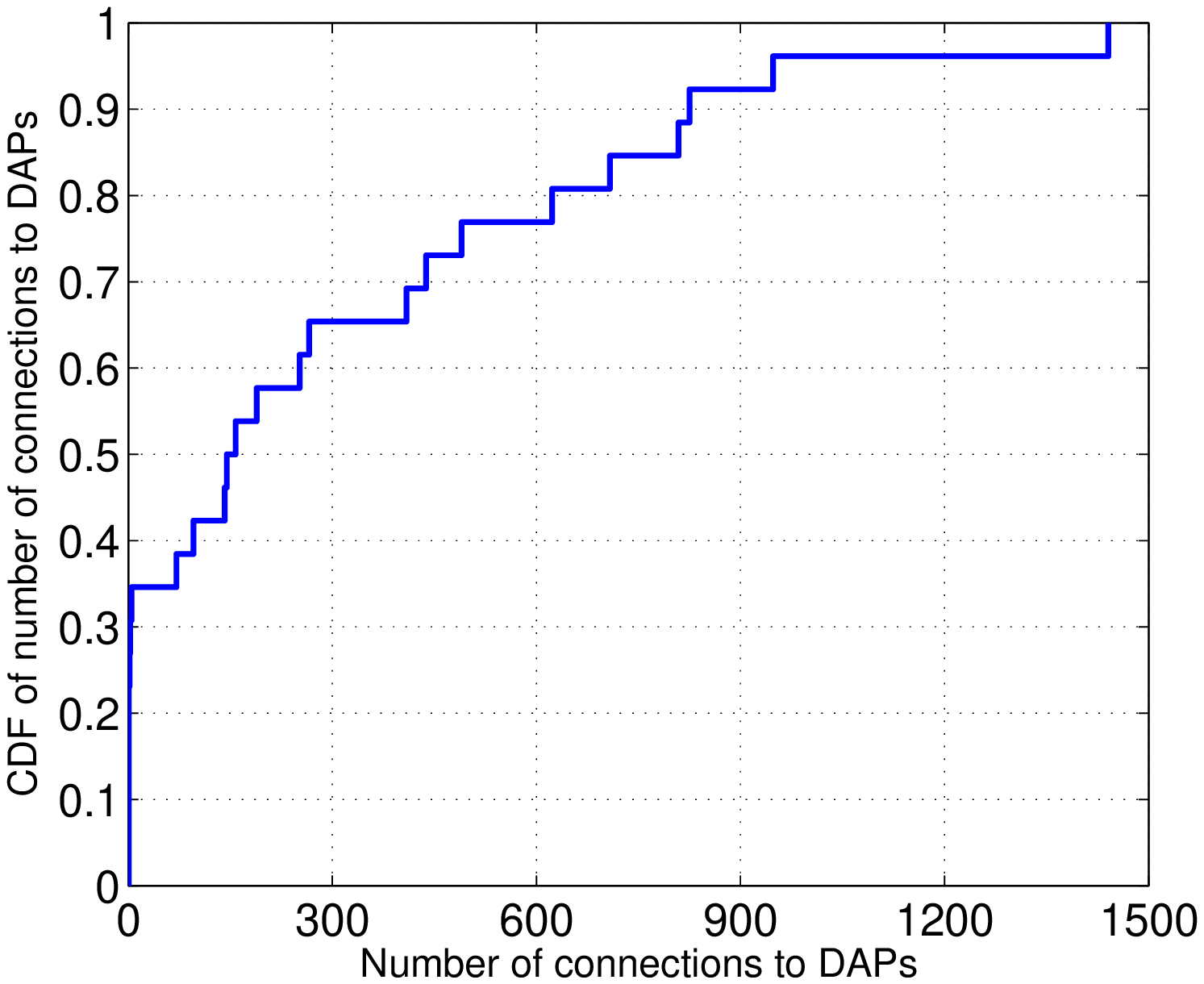}\label{fig:cdf_con}} 
\subfigure[]{\includegraphics[width=0.52\columnwidth]{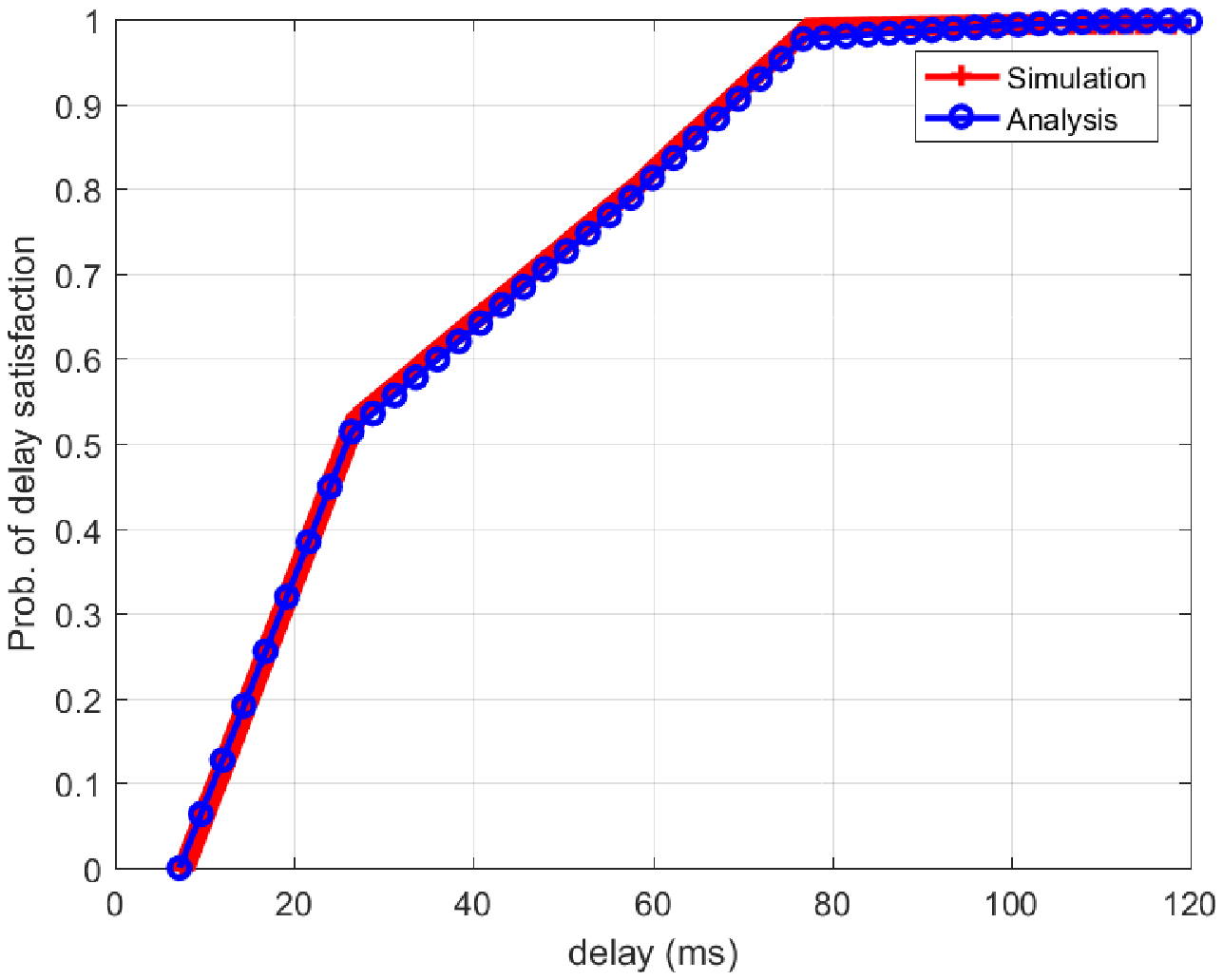}}\caption{ (a) CDF of the total number of connections to DAPs (the mean value is $322$). (b) Comparison of the analysis and simulation for the probability of delay satisfaction as a function of deadline.}
\label{fig:simulation_analysis_comparison_one_hop}
\end{figure}
In order to validate our assumptions and delay model derived in Section~\ref{sec:sys}, we use the network simulator-3 (NS3)~\cite{NS3} software to simulate the SM-to-relay transmissions in the Kamloops scenario. 
Each SM generates packets based on the traffic classes listed in Table~\ref{tab:trafficProperties}. 
We measure the total delay experienced by each packet as the difference between the time it is successfully received by the destination and its generation time. Figure~\ref{fig:simulation_analysis_comparison_one_hop} compares the empirical delay distribution with the analytical probability of delay satisfaction for the packets that have been generated from an SM, which has $124$ feeding nodes and $126$ neighbour nodes. 
Nine of the neighbours have respectively $1244$, $330$, $319$, $233$, $108$, $58$, $53$, $26$, $5$ feeding nodes and the other $117$ nodes do not have any. 
As it can be seen from Figure~\ref{fig:simulation_analysis_comparison_one_hop}, the probability of latency satisfaction obtained from simulations closely matches the values obtained from the analysis in Section~\ref{subsec:lat}. This verifies that the assumptions made in the system model are valid for the traffic classes listed in Table~\ref{tab:trafficProperties}. Specifically, under the mixed traffic model the distribution of packet generations in each SM can be well approximated with a Poisson distribution, and the distribution of packet arrival in the forwarding nodes can be also assumed to follow a Poission distribution.
\vspace*{-5mm}
\subsection{Number of DAPs}
Figure~\ref{fig:daptopology} shows the result of the DAP placement algorithm for the whole Kamloops scenario. In the first iteration of the algorithm, $19$~poles, marked with red squares, are selected for DAP placement such that the network coverage can be ensured. In the next $3$ iterations, 
 $6$~additional poles, marked with cyan squares, are added in order to enforce the required reliability for the SMs that do not satisfy the reliability requirement. These SMs are marked with green circles in the Figure. 
For the $19$ SMs which are located in the same building at location $(-1.0, 0.35)$, there is no connectivity solution, as there is no pole or SM in their connectivity range. 

\subsection{Connections per Pole}

\begin{figure*}[!t]
\hspace*{-1cm}
\subfigure[]{\includegraphics[width=0.41\paperwidth]{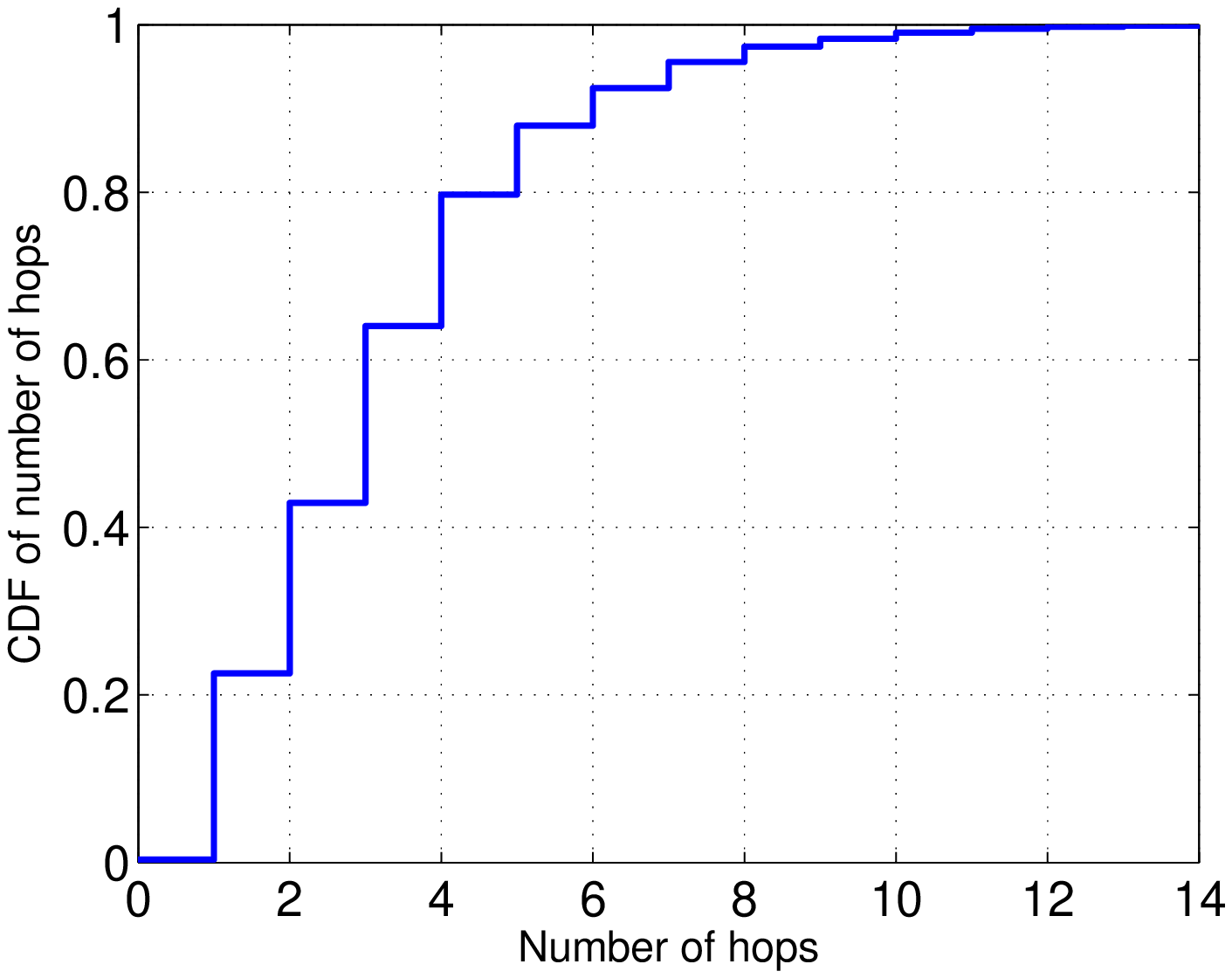}\label{fig:cdf_hops}}
\subfigure[]{\includegraphics[width=0.54\columnwidth]{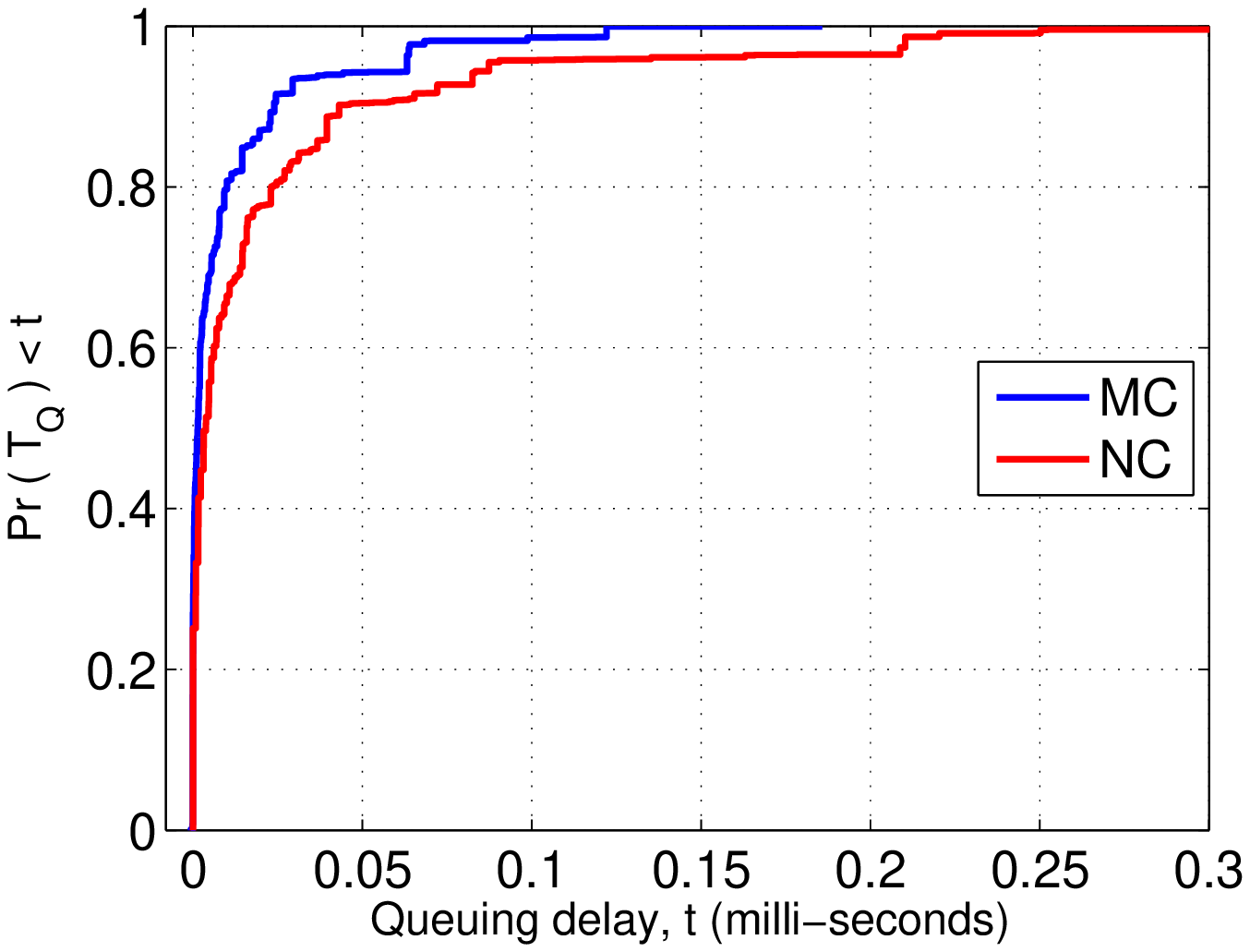}\label{fig:QDCDF}}
\caption{ (a) CDF of the number of hops (the mean value is $3.2$). (b) CDF of queuing delay for the MC and NC traffic.}
\label{fig:cdfs}
\vspace*{-1cm}
\end{figure*}

Figure~\ref{fig:cdf_con} shows the empirical cumulative distribution function (CDF) of the number of connections to the DAPs for the Kamloops scenario. It is observed that around $80\%$ of the DAPs have less than $623$ SM connections. 
We also note that about $35\%$ of the DAPs have less than $5$ connections which is due to the several rural areas with sparse location of smart meters, e.g. for $x < -6.0$ in Figure~\ref{fig:DAP_placement}. To reduce the number of DAPs with few connectivities, the installation of range extenders would be beneficial. 


\subsection{Number of Hops}
Figure~\ref{fig:cdf_hops} shows the distribution of the number of hops for SM-DAP connections in the network for the Kamloops scenario.
As can be seen, around $22\%$ of the nodes are directly connected to DAPs, and 
$90\%$ of the nodes are within a $6$-hop connectivity from a DAP. For the farther nodes, 
our algorithm ensures that their obtained reliability is still within what is required. 
This shows the flexibility of our algorithm compared to~\cite{aoun2006gateway} and~\cite{souza2013optimal}, where they address latency through considering a fixed number of hops, while our algorithm selects the DAP locations and number of hops based on the network topology, SM to SM and SM to pole distances and number of competitors at each hop. The dynamic selection of number of hops based on these parameters makes it possible to access farther SMs with the lowest number of DAPs, without compromising the required latency.

\vspace*{-5mm}
\subsection{Queuing Delay}
Figure~\ref{fig:QDCDF} shows the CDF of the queuing delays observed for the mission-critical and non-critical traffic for the Kamloops scenario. The maximum queuing delay observed for mission-critical traffic is around $0.17$~ms and the maximum queuing delay observed for non-critical traffic is around $0.30$~ms.
The small queuing delay observed is due to the low data rate at the nodes.


%
%
%
%

%% file: Aalamifar_Lampe_arxiv.bbl
\begin{thebibliography}{10}
\providecommand{\url}[1]{#1}
\csname url@samestyle\endcsname
\providecommand{\newblock}{\relax}
\providecommand{\bibinfo}[2]{#2}
\providecommand{\BIBentrySTDinterwordspacing}{\spaceskip=0pt\relax}
\providecommand{\BIBentryALTinterwordstretchfactor}{4}
\providecommand{\BIBentryALTinterwordspacing}{\spaceskip=\fontdimen2\font plus
\BIBentryALTinterwordstretchfactor\fontdimen3\font minus
  \fontdimen4\font\relax}
\providecommand{\BIBforeignlanguage}[2]{{%
\expandafter\ifx\csname l@#1\endcsname\relax
\typeout{** WARNING: IEEEtran.bst: No hyphenation pattern has been}%
\typeout{** loaded for the language `#1'. Using the pattern for}%
\typeout{** the default language instead.}%
\else
\language=\csname l@#1\endcsname
\fi
#2}}
\providecommand{\BIBdecl}{\relax}
\BIBdecl

\bibitem{depuru2011smart}
S.~S. S.~R. Depuru, L.~Wang, and V.~Devabhaktuni, ``Smart meters for power
  grid: Challenges, issues, advantages and status,'' \emph{Renewable and
  Sustainable Energy Reviews}, vol.~15, no.~6, pp. 2736--2742, 2011.

\bibitem{darby2006effectiveness}
S.~Darby, ``The effectiveness of feedback on energy consumption,'' \emph{A
  Review for DEFRA of the Literature on Metering, Billing and direct Displays},
  vol. 486, 2006.

\bibitem{OpenSGForum}
\BIBentryALTinterwordspacing
``{SG Network System Requirements Specification v5.1},'' Open {S}mart {G}rid
  (Open {SG}), Tech. Rep., 2010. [Online]. Available:
  \url{http://osgug.ucaiug.org/default.aspx.}
\BIBentrySTDinterwordspacing

\bibitem{PAP2}
\BIBentryALTinterwordspacing
``{NIST PAP2} guidelines for assessing wireless standards for smart grid
  application,'' 2012. [Online]. Available:
  \url{http://tinyurl.com/NIST-PAP2-Guidelines}
\BIBentrySTDinterwordspacing

\bibitem{mclaughlin2009energy}
S.~McLaughlin, D.~Podkuiko, and P.~McDaniel, ``Energy theft in the advanced
  metering infrastructure,'' in \emph{Critical Information Infrastructures
  Security}.\hskip 1em plus 0.5em minus 0.4em\relax Springer, 2009, pp.
  176--187.

\bibitem{WiGrid}
\BIBentryALTinterwordspacing
{\relax WiMAX Forum}, ``{WiMAX} forum system profile requirements for smart
  grid applications: Requirements for {WiGRID},'' 2013. [Online]. Available:
  \url{http://tinyurl.com/Smart-Grid-Utility-Requirement}
\BIBentrySTDinterwordspacing

\bibitem{atkinson2014leveraging}
G.~Atkinson and M.~Thottan, ``Leveraging advanced metering infrastructure for
  distribution grid asset management,'' in \emph{IEEE Conference on Computer
  Communications Workshops (INFOCOM WKSHPS)}, 2014, pp. 670--675.

\bibitem{bartoli2010secure}
A.~Bartoli, J.~Hern{\'a}ndez-Serrano, M.~Soriano, M.~Dohler, A.~Kountouris, and
  D.~Barthel, ``Secure lossless aggregation for smart grid {M2M} networks,'' in
  \emph{IEEE International Conference on Smart Grid Communications}, 2010, pp.
  333--338.

\bibitem{niyato2012cooperative}
D.~Niyato and P.~Wang, ``Cooperative transmission for meter data collection in
  smart grid,'' \emph{{IEEE} Commun. Mag.}, vol.~50, no.~4, pp. 90--97, 2012.

\bibitem{aoun2006gateway}
B.~Aoun, R.~Boutaba, Y.~Iraqi, and G.~Kenward, ``Gateway placement optimization
  in wireless mesh networks with qos constraints,'' \emph{{IEEE} J. Sel. Areas
  Commun.}, vol.~24, no.~11, pp. 2127--2136, 2006.

\bibitem{chatzigiannakis2006sink}
I.~Chatzigiannakis, A.~Kinalis, and S.~Nikoletseas, ``Sink mobility protocols
  for data collection in wireless sensor networks,'' in \emph{ACM International
  Workshop on Mobility Management and Wireless Access}, 2006, pp. 52--59.

\bibitem{sheng2006data}
B.~Sheng, Q.~Li, and W.~Mao, ``Data storage placement in sensor networks,'' in
  \emph{ACM International Symposium on Mobile Ad Hoc Networking and Computing},
  2006, pp. 344--355.

\bibitem{alsalih2010placement}
W.~Alsalih, H.~Hassanein, and S.~Akl, ``Placement of multiple mobile data
  collectors in wireless sensor networks,'' \emph{Ad Hoc Networks}, vol.~8,
  no.~4, pp. 378--390, 2010.

\bibitem{lin2010optimal}
B.~Lin, P.-H. Ho, L.-L. Xie, X.~Shen, and J.~Tapolcai, ``Optimal relay station
  placement in broadband wireless access networks,'' \emph{{IEEE} Trans. Mobile
  Comput.}, vol.~9, no.~2, pp. 259--269, 2010.

\bibitem{lee2011eqar}
S.~Lee and M.~F. Younis, ``{EQAR}: Effective {QoS}-aware relay node placement
  algorithm for connecting disjoint wireless sensor subnetworks,'' \emph{{IEEE}
  Trans. Comput.}, vol.~60, no.~12, pp. 1772--1787, 2011.

\bibitem{lee2009qos}
S.~Lee and M.~Younis, ``{QoS}-aware relay node placement in a segmented
  wireless sensor network,'' in \emph{IEEE International Conference on
  Communications (ICC)}, 2009, pp. 1--5.

\bibitem{doe2010communications}
``Communications requirements of smart grid technologies,'' US Department of
  Energy, Tech. Rep., 2010.

\bibitem{studio1software}
``{IBM CPLEX Optimization Studio},'' \emph{URL
  https://tinyurl.com/ibm-cplex-optimization-studio}, 2012.

\bibitem{glpk}
\BIBentryALTinterwordspacing
A.~Makhorin, ``{GLPK (GNU linear programming kit)},'' 2016. [Online].
  Available: \url{http://​www.​gnu.​org/​software/​glpk/}
\BIBentrySTDinterwordspacing

\bibitem{souza2013optimal}
G.~Souza, F.~Vieira, C.~Lima, G.~Junior, M.~Castro, and S.~Araujo, ``Optimal
  positioning of {GPRS} concentrators for minimizing node hops in smart grids
  considering routing in mesh networks,'' in \emph{IEEE PES Conference On
  Innovative Smart Grid Technologies Latin America (ISGT LA)}, 2013, pp. 1--7.

\bibitem{rolim2015modelling}
G.~Rolim, D.~Passos, I.~Moraes, and C.~Albuquerque, ``Modelling the data
  aggregator positioning problem in smart grids,'' in \emph{IEEE International
  Conference on Computer and Information Technology (CIT)}, 2015, pp. 632--639.

\bibitem{ali2011set}
K.~Ali, W.~Alsalih, and H.~Hassanein, ``Set-cover approximation algorithms for
  load-aware readers placement in {RFID} networks,'' in \emph{IEEE
  International Conference on Communications (ICC)}, 2011, pp. 1--6.

\bibitem{aalamifar2014cost}
F.~Aalamifar, G.~N. Shirazi, M.~Noori, and L.~Lampe, ``Cost-efficient data
  aggregation point placement for advanced metering infrastructure,'' in
  \emph{IEEE International Conference on Smart Grid Communications}, 2014, pp.
  344--349.

\bibitem{NANkong2015}
P.~Y. Kong, ``Wireless neighborhood area networks with {QoS} support for demand
  response in smart grid,'' \emph{{IEEE} Trans. Smart Grid}, vol.~7, no.~4, pp.
  1913--1923, July 2016.

\bibitem{niyato2011machine}
D.~Niyato, L.~Xiao, and P.~Wang, ``Machine-to-machine communications for home
  energy management system in smart grid,'' \emph{{IEEE} Commun. Mag.},
  vol.~49, no.~4, pp. 53--59, 2011.

\bibitem{802154gSUN2012}
``{IEEE Standard for Local and metropolitan area networks--Part 15.4: Low-Rate
  Wireless Personal Area Networks (LR-WPANs) Amendment 3: Physical Layer (PHY)
  Specifications for Low-Data-Rate, Wireless, Smart Metering Utility
  Networks},'' \emph{IEEE Std 802.15.4g-2012 (Amendment to IEEE Std
  802.15.4-2011)}, pp. 1--252, April 2012.

\bibitem{OSG}
\BIBentryALTinterwordspacing
``Open {S}mart {G}rid (open {SG}).'' [Online]. Available:
  \url{http://osgug.ucaiug.org/default.aspx.}
\BIBentrySTDinterwordspacing

\bibitem{mohassel2014survey}
R.~R. Mohassel, A.~Fung, F.~Mohammadi, and K.~Raahemifar, ``A survey on
  advanced metering infrastructure,'' \emph{International Journal of Electrical
  Power \& Energy Systems}, vol.~63, pp. 473--484, 2014.

\bibitem{GomezLeasedLine2013}
F.~Gomez-Cuba, R.~Asorey-Cacheda, and F.~Gonzalez-Castano, ``Smart grid
  last-mile communications model and its application to the study of leased
  broadband wired-access,'' \emph{{IEEE} Trans. Smart Grid}, vol.~4, no.~1, pp.
  5--12, Mar 2013.

\bibitem{PAP2014}
\BIBentryALTinterwordspacing
{Participants of the Priority Action Plan 2 working group}, ``{NIST} smart grid
  interoperability panel {PAP2} guidelines for assessing wireless standards for
  smart grid application,'' 2014. [Online]. Available:
  \url{http://nvlpubs.nist.gov/nistpubs/ir/2014/NIST.IR.7761r1.pdf}
\BIBentrySTDinterwordspacing

\bibitem{shi2013analytical}
Z.~Shi, C.~Beard, and K.~Mitchell, ``Analytical models for understanding space,
  backoff, and flow correlation in {CSMA} wireless networks,'' \emph{Wireless
  networks}, vol.~19, no.~3, pp. 393--409, 2013.

\bibitem{Marco2012analytical}
P.~Di~Marco, P.~Park, C.~Fischione, and K.~H. Johansson, ``Analytical modeling
  of multi-hop {IEEE 802.15.4} networks,'' \emph{{IEEE} Trans. Veh. Technol.},
  vol.~61, no.~7, pp. 3191--3208, 2012.

\bibitem{ray2005performance}
S.~Ray, D.~Starobinski, and J.~B. Carruthers, ``Performance of wireless
  networks with hidden nodes: A queuing-theoretic analysis,'' \emph{Computer
  Communications}, vol.~28, no.~10, pp. 1179--1192, 2005.

\bibitem{shi2006starvation}
J.~Shi, T.~Salonidis, and E.~W. Knightly, ``{Starvation mitigation through
  multi-channel coordination in {CSMA} multi-hop wireless networks},'' in
  \emph{ACM International Symposium on Mobile Ad Hoc Networking and Computing},
  2006, pp. 214--225.

\bibitem{bose2013introduction2q}
S.~K. Bose, \emph{An introduction to queueing systems}.\hskip 1em plus 0.5em
  minus 0.4em\relax Springer Science \& Business Media, 2013.

\bibitem{gao2009new}
J.~Gao, J.~Hu, and G.~Min, ``{A new analytical model for slotted IEEE 802.15. 4
  medium access control protocol in sensor networks},'' in \emph{International
  Conference on Communications and Mobile Computing}, vol.~2.\hskip 1em plus
  0.5em minus 0.4em\relax IEEE, 2009, pp. 427--431.

\bibitem{misic2006performance}
J.~Misic, S.~Shafi, and V.~B. Misic, ``{Performance of a beacon enabled IEEE
  802.15.4 cluster with downlink and uplink traffic},'' \emph{{IEEE} Trans.
  Parallel Distrib. Syst.}, vol.~17, no.~4, pp. 361--376, 2006.

\bibitem{baz2015analysis}
M.~Baz, P.~D. Mitchell, and D.~A. Pearce, ``Analysis of queuing delay and
  medium access distribution over wireless multihop {PANs},'' \emph{{IEEE}
  Trans. Veh. Technol.}, vol.~64, no.~7, pp. 2972--2990, 2015.

\bibitem{Fernandez_closedform}
M.~Fernandez and S.~Williams, ``Closed-form expression for the
  {Poisson-Binomial} probability density function,'' \emph{{IEEE} Trans.
  Aerosp. Electron. Syst.}, vol.~46, no.~2, pp. 803--817, April 2010.

\bibitem{Bertsekas1992DataNet}
D.~Bertsekas and R.~Gallager, \emph{Data Networks (2Nd Ed.)}.\hskip 1em plus
  0.5em minus 0.4em\relax Upper Saddle River, NJ, USA: Prentice-Hall, Inc.,
  1992.

\bibitem{meng2014smart}
W.~Meng, R.~Ma, and H.-H. Chen, ``Smart grid neighborhood area networks: a
  survey,'' \emph{{IEEE} Netw.}, vol.~28, no.~1, pp. 24--32, 2014.

\bibitem{alsalih2008placement}
W.~Alsalih, S.~Akl, and H.~Hassanein, ``Placement of multiple mobile data
  collectors in underwater acoustic sensor networks,'' in \emph{IEEE
  International Conference on Communications (ICC)}, 2008, pp. 2113--2118.

\bibitem{samet1990design}
H.~Samet, \emph{The design and analysis of spatial data structures}.\hskip 1em
  plus 0.5em minus 0.4em\relax Addison-Wesley Reading, MA, 1990, vol. 199.

\bibitem{chandra2004optimizing}
R.~Chandra, L.~Qiu, K.~Jain, and M.~Mahdian, ``Optimizing the placement of
  internet taps in wireless neighborhood networks,'' in \emph{IEEE
  International Conference on Network Protocols}, 2004, pp. 271--282.

\bibitem{Feige1998SetCover}
U.~Feige, ``{A Threshold of Ln N for Approximating Set Cover},'' \emph{{Journal
  of the ACM}}, vol.~45, no.~4, pp. 634--652, Jul. 1998.

\bibitem{zangwill1969nonlinear}
W.~I. Zangwill, \emph{Nonlinear programming: {A} unified approach}.\hskip 1em
  plus 0.5em minus 0.4em\relax Prentice-Hall, 1969.

\bibitem{luenberger1973introduction}
D.~G. Luenberger, \emph{Introduction to linear and nonlinear
  programming}.\hskip 1em plus 0.5em minus 0.4em\relax Addison-Wesley Reading,
  MA, 1973, vol.~28.

\bibitem{NS3}
\BIBentryALTinterwordspacing
``{Network Simulator-3 (NS-3)},'' 2015. [Online]. Available:
  \url{http://www.nsnam.org/}
\BIBentrySTDinterwordspacing

\bibitem{CPLEXHelp}
\BIBentryALTinterwordspacing
``Complexity of the integer linear programming,'' 2014. [Online]. Available:
  \url{https://www.ibm.com/developerworks/community/forums/html/topic?id=77777777-0000-0000-0000-000014393637}
\BIBentrySTDinterwordspacing

\bibitem{ResearchGate}
\BIBentryALTinterwordspacing
``How to measure the difficulty of a mixed-linear integer programming ({MILP})
  problem,'' 2015. [Online]. Available:
  \url{https://www.researchgate.net/post/How_to_measure_the_difficulty_of_a_Mixed-Linear_Integer_Programming_MILP_problem}
\BIBentrySTDinterwordspacing

\bibitem{CPLEXForum}
\BIBentryALTinterwordspacing
``Guidelines for estimating {CPLEX} memory requirements based on problem
  size,'' 2012. [Online]. Available:
  \url{http://www-01.ibm.com/support/docview.wss?uid=swg21399933}
\BIBentrySTDinterwordspacing

\end{thebibliography}
